\documentclass{article}
\usepackage{amssymb}
\usepackage{epsfig}
\usepackage{amsmath}

\input{tcilatex}

\begin{document}


\title{\textbf{General Relativity, the Cosmological Constant and 
Modular Forms.}}
\author{G. V. Kraniotis 
\footnote{Present address: Martin-Luther-Universit$\rm{\ddot a}$t
  Halle-Wittenberg, Fachgruppe Theoretische Physik, Friedmann-Bach-Platz 6,
06099 Halle, Germany, kraniotis@physik.uni-halle.de} 
\\
Centre for Theoretical Physics, \\
\footnote{g.kraniotis@sussex.ac.uk, SUSX-TH/01-018} 
University of Sussex, \\
Brighton, BN1 9QJ \and S. B. Whitehouse  \\
Royal Holloway, University of London, 
\footnote{sbwphysics@aol.com,RHCPP01-04T} \\
Physics Department, \\
Egham Surrey TW20-0EX, U.K.}
\maketitle

\begin{abstract}
Strong field (exact) solutions of the gravitational field equations of
General Relativity in the presence of a Cosmological Constant are
investigated. In particular, a full exact solution is derived within the
inhomogeneous Szekeres-Szafron family of space-time line element with a
nonzero Cosmological Constant. The resulting solution connects, in an
intrinsic way, General Relativity with the theory of modular forms and
elliptic curves. The homogeneous
FLRW limit of the above space-time elements is recovered and we solve
exactly the resulting Friedmann Robertson field equation with the
appropriate matter density for generic values of the Cosmological Constant $%
\Lambda $ and curvature constant $K$. A formal expression for the
Hubble constant is derived. The cosmological implications of the resulting
non-linear solutions are systematically investigated. Two particularly
interesting solutions i) the case of a flat universe $K=0,\Lambda \not=0$
and ii) a case with all three cosmological parameters non-zero, are
described by elliptic curves with the property of complex multiplication
and absolute modular invariant $j=0$ and $1728$, respectively. The 
possibility that
all non-linear solutions of General Relativity are expressed in terms of
theta functions associated with Riemann-surfaces is discussed.
\end{abstract}

\bigskip

\bigskip \newpage

\section{ Introduction}

\subsection{\protect\bigskip Motivation}

The Cosmological Constant $\Lambda $ \cite{ALVERTOS} is presently  
at the epicentre of
contemporary physics \cite{GVKSBW} and plays a significant role in fields,
such as: cosmology, astronomy, General Relativity, particle physics, and 
string theory.

Experimental evidence has been growing suggesting that the Cosmological
Constant has a small but non-zero value \cite{PERLMUTTER},\cite{FELTENGER},%
\cite{CARROLL}. In particular, the results from high redshift 
Type Ia supernovae
observations suggest an accelerating Universe (positive Cosmological
Constant) where the age of the Universe and Hubble's Constant $H$ \ are
approximately, $\sim $14.5 billion years and $70$ $Kms^{-1}Mpc^{-1}$,
respectively. Other indirect experimental evidence (Turner \cite{TURNER},
Bahcall et al.\cite{BAHCALL}) has put bounds on the energy densities ratio's
of matter, dark energy and curvature, given by: $\Omega _{m}\equiv 8\pi
G_{N}\rho _{matter}/(3H^{2})\sim 0.3,$ $\Omega _{\Lambda }\equiv \frac{%
c^{2}\Lambda }{3H^{2}}\sim 0.6-0.7,$ $\Omega _{K}\equiv
-Kc^{2}/(R(t)H)^{2}\sim 0$, respectively. In these definitions $G_{N}$ is
Newton's constant, $\rho _{matter}$ denotes the density of matter, $K$ the
curvature coefficient and $R(t)$ the scale factor.

We note here that The Cosmological Constant has been identified with a
``dark'' exotic form of energy that is smoothly distributed and which
contributes 60 -70\% to the critical density of the Universe \cite{TURNER}
(See Cohn \cite{COHN} and Carroll and Press \cite{CARROLL} for a pedagogical
review of the Cosmological Constant and Turner \cite{TURNER} for a review of
Dark Matter and Dark Energy).

Recent data  from the Boomerang experiment \cite{DEBERNAR} provide a firm 
evidence for the case of flat Universe, $\Omega _{K}\sim 0$.

It was suggested that the effects of the Cosmological Constant could be felt
on a Galactic scale \cite{GVKSBW} which would lead to the modification of
Newton's inverse square law. This approach was used in order to try and
explain the behaviour of the Galactic velocity rotation curves. The 
dynamical effects
of $\Lambda $ at megaparcec scale were investigated by Axenides et al \cite%
{MAEFLP}.

Some of the theoretical justifications for a non-zero Cosmological Constant
are discussed by Overduin et al \cite{FELTENGER}. 

The above arguments based on observations motivate a program of deriving 
exact solutions of General Relativity with a non-zero Cosmological Constant 
$\Lambda$ and eventually apply these solutions to problems in cosmology  
and investigate dynamical effects of $\Lambda$ at various astrophysical scale.

The main objectives of this paper are: first to derive new exact solutions of General
Relativity, for a non-zero Cosmological Constant, in the 
context of the inhomogeneous Szekeres-Szafron family of 
space-time line elements, and discuss at a preliminary level the effect 
of inhomogeneity in cosmology. Secondly, to present in 
a systematic manner the cosmological implications of the homogeneous Friedmann, Robertson, Lemaitre, Walker (FRLW) limit of the 
above inhomogeneous metric. Details of the cosmological implications of the 
exact inhomogeneous solution will appear elsewhere. 
The material of this paper is organized as follows.
In section 2 we derive a full exact solution of the gravitational field equations with a non-zero Cosmological Constant in the inhomogeneous class of 
Szekeres-Szafron models. The solution in a closed analytic form is expressed by the 
Weierstra$\ss$ elliptic function. Our motivation for studying this particular class of inhomogeneous models is twofold. First 
the solution described above is new, and in general  
the inhomogenous cosmology is 
only a partially explored territory. Also the 
Szekeres-Szafrom metric is fairly general and consequently our solution of 
the resulting field equations represents the most general exact solution of General Relativity obtained to date.
We note that the lack of general exact solutions of the equations of 
General Relativity has been an obstacle in obtaining the full realistic 
implications of the theory for cosmology.

Secondly, a homogeneous Friedmann, Robertson, Lemaitre, Walker limit can be derived from this family and therefore a comparison 
between inhomogeneous and homogeneous cosmology can in principle 
be made. However, in this paper we do not attempt any such comparison. 
This will be a subject of a future publication. We discuss however, at a 
preliminary level where we expect the effect of inhomogeneity to appear.
In section 3 we derive the FRLW limit and solve the Friedmann equation 
for the scale factor in terms of Weierstra$\ss$sche modular forms using the techniques of section 2. A formal 
expression for the Hubble's Constant is also derived.  Section 4 provides 
the necessary mathematical background for the use of the 
elliptic functions and elliptic curves in the solutions obtained. In section 5 we systematically investigate the cosmological 
implications of the exact solution of the  Friedmann equation. 
The last section is used for our conclusions.
We also include in appendices some of the formal calculations that were 
performed for this paper, and 
we emphasize that the same techniques are also usefull for the solution
of other important non-linear partial differential equations of 
mathematical physics.
The present work is expected to be of interest to a wide audience, ranging 
from observational astronomers, relativitists, to theorists and applied
mathematicians. We have therefore tried to point the reader to the appropriate
literature for the necessary physics and mathematics background, and kept 
the mathematical formalism  to the appropriate  level for making the
presentation self-contained.
\bigskip

\bigskip

\subsection{Inhomogeneous versus Homogeneous Cosmologies}

As already mentioned in the introduction one of the reasons for choosing 
to  study inhomogeneous cosmologies was to 
explore the relatively unknown territory of inhomogeneous exact solutions 
of Einstein's field equations. The Friedmann, Lemaitre, Robertson, Walker (FLRW) models are a realistic first approximation to the geometry 
and physics of our Universe. By deriving and investigating exact 
solutions of the field 
equations of General Relativity with general inhomogeneoous metric elements 
the theory will be prepared to meet any challenges posed by an inhomogeneous 
Universe. As it is emphasized by Krasinski in \cite{KRASGR} inhomogeneous 
solutions are very convenient in considering problems such as 
the anisotropies 
in the cosmic microwave background radiation generated by 
density inhomogeneities in the Universe, evolutions of voids, the issue of 
singularities, possible effects of the expansion of the Universe on 
planetary orbits and others.


Together the books, ''Exact Solutions of Einstein's Field Equations'' \cite%
{exact}, and ''Inhomogeneous Cosmological Models'' \cite{KRASINSKI},
including references within, provide a comprehensive review of the presently
known families of exact solutions to Einstein's Field equations.

The Szekeres-Szafron family of solutions were chosen for study as they
represent a well-known inhomogeneous cosmological model, whose metric is
given by:

\begin{equation}
ds^{2}=dt^{2}-e^{2\beta \left( x,y,z,t\right) }(dx^{2}+dy^{2})-e^{2\alpha
\left( x,y,z,t\right) }dz^{2}.  \label{metriki}
\end{equation}
where the functions $\alpha
\left( x,y,z,t\right),\beta\left( x,y,z,t\right)$ are to be determined by 
the field equations. 

The characteristics, properties and formal solutions
for the energy densities, for this family of exact inhomogeneous solutions
has been collated and documented by Krasinski \cite{KRASINSKI}.

The Einstein equations with a non-zero Cosmological Constant, and a perfect
fluid source energy momentum tensor $T_{\mu \nu }$ have the form, 

\begin{equation}
G_{\mu \nu }:=R_{\mu\nu}-\frac{1}{2}g_{\mu\nu}R=-(\Lambda g_{\mu \nu }+\kappa 
T_{\mu \nu }),  \label{INHOGRE}
\end{equation}
with
\begin{equation}
T_{\mu \nu }=(\tilde{P}+\tilde{\rho})u_{\mu }u_{\nu }-\tilde{P}g_{\mu \nu },\quad u_{\mu }=\delta
_{\mu }^{0},
\label{fluid}
\end{equation}
here $u_{\mu}$ , $\tilde{P}$ and $\tilde{\rho} $ are the four-velocity, 
pressure and fluid energy density respectively. 
In this paper we study the model with Cosmological 
constant and  pressureless matter ($\tilde{P}=0$) in Eq.\ref{fluid}. 
To the best of our knowledge the
exact solution of the non-linear partial differential field equations (\ref%
{INHOGRE}) resulting from the metric Eq.(\ref{metriki}), presented in 
this paper is new and represents the first solution in a closed analytic form. Here we
acknowledge the paper by Barrow and Stein-Schabes \cite{BARROW} who were the
first to derive a solution which was defined as a formal integral. In
particular, as we shall see $\alpha (x,y,z,t)$, $\beta (x,y,z,t)$ and the
energy density are expressed in terms of the Weierstra\ss 's modular forms.
Consequently, our exact
solutions for the inhomogeneous Szekeres-Szafron space-time element with a
non-zero $\Lambda $ lie on an elliptic curve.



\section{Exact Solution of the Gravitational Field Equations for the
Szekeres-Szafron Line Element}

\subsection{Approach, Definitions and Assumptions}

This section will outline the approach, definition and assumptions required
to derive the general expression for $\alpha (x,y,z,t)$ and $\beta (x,y,z,t)$
from the Field Equations. The following sections will describe the
determination of the Einstein's Field Equations in the Szafron form, and the
derivation of the exact solution for the line element.

We start by assuming that there exist co-ordinates which support the metric
line element Eq.(\ref{metriki}).







The solutions are classified as either type $\beta ^{\prime }=0$ or $\beta
^{\prime }\neq 0$, where $\beta ^{\prime }=\partial \beta /\partial z,$
depending on what initial trial solution is chosen. In this section the more
general $\beta ^{\prime }\neq 0$ type will be studied \cite{KRASINSKI}.

\subsection{Einstein and Szafron Form of the Field Equations}

Determining the Einstein field equations from the metric is straight forward
but tedious. Expressions for the non-zero components of the Ricci $\ $tensor 
$R_{\mu \nu ,}$ and scalar $R$: $R_{00},$ $R_{11,}$ $R_{22},$ $R_{33},$ $R$
are given in appendix A for completeness.

The original solution for this metric was given by Szafron \cite{SZAFRON},
his starting point was to introduce a pair of complex variables given by:

\begin{equation}
\xi :=x+i\ y,\quad \overline{\xi }:=x-i\ y.
\end{equation}

Following his approach we re-derived the field equations which are given
below (due to an ordering difference of terms
in the metric, our subscript three is equivalent to his one, i.e. $%
G_{11}^S\equiv G_{33}).$ The field equations are:

\begin{equation}
G_{0}^{0}-2G_{\xi }^{\xi }-G_{3}^{3}=2\ (\ddot{\alpha}+2\ddot{\beta}+\dot{%
\alpha}^{2}+2\dot{\beta}^{2})=-\kappa \left( \rho +3P\right)=
-(-2\Lambda+\kappa \tilde{\rho}),
\label{EINSTEIN1}
\end{equation}

\begin{equation}
\frac{1}{2}G_{3}^{0}=\dot{\beta}^{\prime }-\beta ^{\prime }(\dot{\alpha}-%
\dot{\beta})=0,  \label{ALBERT}
\end{equation}

\begin{equation}
G_{3}^{3}=4e^{-2\beta }\beta _{\xi \overline{\xi }}+e^{-2\alpha }\beta
^{^{\prime }2}-2\ddot{\beta}-3\dot{\beta}^{2}=-(-\kappa P)=
-\Lambda,  \label{FIELD}
\end{equation}

\begin{equation}
G_{\xi }^{\xi }=-e^{-2\alpha }\left( -\beta ^{\prime \prime }+\alpha
^{\prime }\beta ^{\prime }-\beta ^{\prime 2}\right) +2e^{-2\beta }(\alpha
_{\xi \overline{\xi }}+\alpha _{\xi }\alpha _{\overline{\xi }})-(\ddot{\alpha%
}+\ddot{\beta}+\dot{\alpha}^{2}+\dot{\beta}^{2}+\dot{\alpha}\dot{\beta}%
)=-(-\kappa P)=-\Lambda,  \label{FIELD2}
\end{equation}

\begin{equation}
G_{0}^{\xi }=\dot{\alpha}_{\xi }+\dot{\beta}_{\xi }-\alpha _{\xi }(\dot{\beta%
}-\dot{\alpha})=0,  \label{FIELD3}
\end{equation}

\begin{equation}
e^{2\alpha }G_{\xi }^{3}=-\beta _{\xi }^{\prime }+\beta ^{\prime }\alpha
_{\xi }=0,  \label{FIELD6}
\end{equation}

\begin{equation}
\frac{1}{2}e^{2\beta }G_{\xi }^{\overline{\xi }}=\alpha _{\xi \xi }+(\alpha
_{\xi })^{2}-2\beta _{\xi }\alpha _{\xi }=0,  \label{ALVERTOS}
\end{equation}
where $^{\prime }=\partial /\partial z,\ $and $^{.}=\partial /\partial t$,
and
\begin{equation}
\alpha _{\xi }:= \partial \alpha /\partial \xi \equiv \frac{1}{2}%
(\partial \alpha /\partial x-i\ \partial \alpha /\partial y),\quad 
\alpha_{\overline{\xi}} := \partial \alpha /\partial \overline{\xi }=
\frac{1}{2}(\partial \alpha /\partial x+i\ \partial \alpha /\partial y).
\end{equation}
Also we redefined the energy density and pressure by including the 
Cosmological Constant in the energy-momentum tensor:

\begin{eqnarray}
\kappa \rho & := & \kappa \tilde{\rho} + \Lambda   , \nonumber \\
-\kappa P & := & \Lambda.
\label{redefinition}
\end{eqnarray}

\subsection{Exact Solution of the  Szekeres-Szafron Line Element, $\ \protect%
\beta ^{\prime }\neq 0$}

Extending the work of Szekeres \cite{SZEKERES}, Szafron \cite{SZAFRON}
proved that $\dot{\beta}_{\xi }=0,$ which allowed him to propose a trial
solution for the field equations of the form:

\begin{equation}
\beta =\text{Log}\ [\Phi (t,z)]+\nu (z,\xi ,\overline{\ \xi }).
\end{equation}

Since $\beta^{\prime}\neq 0$, Eq.(\ref{ALBERT}) can be integrated to 
obtain:
\begin{equation}
\alpha =\text{Log}[h(z,\xi ,\bar{\xi})\Phi ^{\prime }(t,z)+h(z,\xi ,\bar{\xi}%
)\Phi (t,z)\nu ^{\prime }(z,\xi ,\overline{\xi })].
\end{equation}

It is then straightforward to show that the trial solution identically 
satisfies the rest of the equations (\ref{EINSTEIN1}) - (\ref{ALVERTOS}) with $h=h(z)$, \ leading to
two defining equations given by,

\begin{equation}
e^{-\nu }=A(z)\xi \overline{\xi }+B(z)\xi +\overline{B}(z)\overline{\xi }%
+C(z),
\end{equation}

\begin{equation}
2\frac{\ddot{\Phi}}{\Phi }+\frac{\dot{\Phi}^{2}}{\Phi ^{2}}+\kappa P+\frac{%
K(z)}{\Phi ^{2}}=0,  \label{elliptic}
\end{equation}
where $A(z),B(z),C(z),h(z),$ are arbitrary functions 
($A(z),C(z)$ are real functions, while $B(z)$ is complex), $P$ 
denotes the pressure  and $K(z)$ 
is an arbitrary real function determined by:\footnote{Szafron used $h(z)=1$}

\begin{equation}
A(z)C(z)-B(z)\overline{B}(z)=\frac{1}{4}[h^{-2}(z)+K(z)].
\end{equation}

\bigskip Equation (\ref{elliptic}) can be integrated to give \cite{KRASINSKI}%
:

\begin{equation}
\dot{\Phi}^{2}=-K(z)+2M(z)/\Phi -\frac{1}{3}\frac{\kappa }{\Phi }\int P\frac{%
\partial \Phi ^{3}}{\partial t}dt.
\end{equation}

The solution for $\Phi $ gives us $\alpha \left( x,y,z,t\right) ,\beta
\left( x,y,z,t\right) $ and the general expression for the metric. In this 
paper by putting $\Lambda
=-\kappa P$, see Eq.(\ref{redefinition}), and performing the integration in the above equation, we get the
following non-linear partial differential equation:

\begin{equation}
\dot{\Phi}^{2}=-K(z)+2M(z)/\Phi +\frac{1}{3}\Lambda \Phi ^{2}.
\label{modular}
\end{equation}
Equation (\ref{modular}) 
resembles the equation of an elliptic curve (see section 4) 
and therefore can be integrated parametrically by the use of Weierstra\ss
elliptic functions \footnote{%
For other examples of elliptic integrals see \cite{WHITAKKER}}(see appendix
B).
The solution for $\Phi $ is: 
\begin{equation}
\Phi (t,z)=\frac{M(z)/2}{\wp (u+\epsilon )-\wp (v_{0})},\quad \text{where }%
\wp (v_{0}):=-\frac{K(z)}{12},
\label{solgen}
\end{equation}
where the Weierstra\ss\ invariants are given by:

\begin{equation}
g_{2}=\frac{K(z)^{2}}{12},\quad g_{3}=\frac{1}{216}K(z)^{3}-\frac{1}{12}%
\Lambda M^{2}(z),
\end{equation}
and the time $t$ is obtained from $t=\int \Phi du$ with the result:

\begin{equation}
t+f(z)=\frac{1}{\wp ^{\prime }(v_{0})}\left[ \log \frac{\sigma (u+\epsilon
-v_{0})}{\sigma (u+\epsilon +v_{0})}+2\ (u+\epsilon )\ \zeta (v_{0})\right]
\left( \frac{M(z)}{2}\right) ,
\end{equation}
or alternatively

\begin{equation}
t+f(z)=\sqrt{\frac{3}{\Lambda }}\left[ \log \frac{\sigma (u+\epsilon -v_{0})%
}{\sigma (u+\epsilon +v_{0})}+2\ (u+\epsilon )\ \zeta (v_{0})\right].
\end{equation}
Here $\wp (z|\omega ,\omega ^{\prime })$ , $\zeta (z)=\zeta (z|\omega
,\omega ^{\prime })$,
and $\sigma (z)=\sigma
(z|\omega ,\omega ^{\prime })$ denote the Weierstra\ss ' family of
functions, Weierstra\ss, and Weierstra\ss\ zeta and sigma functions
respectively. Also $\epsilon $ is a constant of integration and $f(z)$ is an
arbitrary function of $z$. The definitions and properties of these functions
will be discussed in detail in section 4. The matter/energy density is 
obtained by the first field equation (\ref{EINSTEIN1}).
We also calculated $\wp^{\prime}(v_0)$ from the defining equation for the 
elliptic curve
\begin{equation}
(\wp^{\prime}(v_0))^2=4\wp^3 (v_0)-g_2 \wp(v_0)-g_3
\label{modulwei}
\end{equation}
substituting $\wp(v_0)=-\frac{K(z)}{12}$ in (\ref{modulwei}) we can show that 
that $(\wp^{\prime}(v_0))^2=\frac{1}{12}\Lambda M^2(z)$, 
and (\ref{solgen}) can be rewritten as: $\Phi(t,z)= \sqrt{\frac{\Lambda}{3}}
\frac{\wp^{\prime}(v_0)}{\wp(u+\epsilon)-\wp(v_0)}$.

\bigskip Substituting our expressions for $\alpha \left( x,y,z,t\right) $
and $\beta \left( x,y,z,t\right) $, given by:

\begin{eqnarray}
e^{\beta } &=&\Phi (t,z)e^{\nu (z,x,y)}, \label{ebeta}\ \qquad \\
e^{\alpha } &=&h(z)e^{-\nu (x,y,z)}\left( e^{\beta }\right) _{,z}  \nonumber \\
&=&h(z)e^{-\nu (x,y,z)}e^{\beta }\left[ \frac{\partial Log(\Phi (t,z))}{%
\partial z}+\frac{\partial \nu }{\partial z}\right] \label{ealpha} ,
\end{eqnarray}
into the metric, Eq.(\ref{metriki}) the final solution becomes:

\begin{eqnarray}
ds^{2}&=&dt^{2}-\frac{(M(z)/2)^{2}}{(\wp (u+\epsilon )-\wp (v_{0}))^{2}}e^{2\nu
}(dx^{2}+dy^{2})-\frac{h^{2}(z)(M(z)/2)^{2}}{(\wp (u+\epsilon )-\wp
(v_{0}))^{2}}(\frac{\partial Log(\Phi (t,z))}{\partial z}+\frac{\partial \nu
(x,y,z)}{\partial z})^{2}dz^{2} \nonumber \\
&=& dt^{2}-\frac{(M(z)/2)^{2}}{(\wp (u+\epsilon )-\wp (v_{0}))^{2}}e^{2\nu
}(dx^{2}+dy^{2})-h^2(z) (\Phi^{\prime}+\Phi \nu^{\prime})^2 dz^2.
\end{eqnarray}

The above solution represents the most general exact solution of inhomogeneous relativistic cosmology obtained to date.


\subsubsection{\protect\bigskip Solution for the Matter Density in terms of
the Cosmological Constant}

In this subsection we derive the expression of the energy density $\rho $
that constitutes a solution of the Einstein field equations within the
Szekeres-Szafron family of space-time elements.

The mass density $\rho $, is obtained from the first field equation (see Eq.
(\ref{EINSTEIN1})),

\begin{equation}
G_{0}^{0}-2G_{\xi }^{\xi }-G_{3}^{3}=-\kappa (\rho +3P),
\end{equation}
Substituting the solution of $\alpha ,\beta $ in Eq.(\ref{EINSTEIN1}) (see
Appendix C for details) the general expression for the energy density $\rho $
is given by:

\begin{equation}
\kappa \rho =\frac{1}{\Phi _{,z}+\Phi \nu _{,z}}\left\{ \frac{2M_{,z}}{\Phi
^{2}}-\frac{1}{3}\kappa \Phi ^{-2}\int P\left( \frac{\partial ^{2}\Phi ^{3}}{%
\partial t\partial z}\right) dt+\frac{6M\nu _{,z}}{\Phi ^{2}}-\kappa \Phi
^{-2}\nu _{,z}\int P\left( \frac{\partial \Phi ^{3}}{\partial t}\right)
dt\right\} .
\end{equation}

Redefining the energy density as in Eq.(\ref{redefinition}),
the final expression for the matter density is given by:

\begin{equation}
\kappa \widetilde{\rho }=\frac{2M_{,z}+6M\nu _{,z}}{\Phi ^{2}(\Phi
_{,z}+\Phi \nu _{,z})}.  \label{DENSITYM}
\end{equation}%

The Friedmann, Lemaitre, Robertson, Walker (FLRW) limit of the above
expression will be discussed in the following section. Here we will only
state that in this limit $\kappa \tilde{\rho}\propto \frac{1}{R(t)^{3}},$
where $R(t)$ is the standard scale factor in the FLRW line-element.

We now discuss at a preliminary level where we expect the effect of inhomogeneity to be important.
One of the areas where the exact inhomogeneous solution of the field 
equations of General Relativity derived in this paper may play a vital role 
is in the field of gravitational lensing.
In particular as is discussed in some length in \cite{EHLERS} light 
propagation through an inhomogeneous universe {\it differs} from that 
through a homogeneous one. Light propagating in an inhomogeneous universe 
is deflected due to gravitational fields, thus limiting the accuracy 
with which positions of distant sources can be determined. 
The basic equations governing the change of cross-sections and shapes 
of ray bundles had been discussed by R.K. Sachs \cite{SA}. There it was 
shown how light propagating through a region of space where the density 
is lower than the average density of the universe diverge relative  
to corresponding light bundles in a homogeneous universe of  the same 
mean matter density \cite{Zel},\cite{DA651}, \cite{GU67}.
R. Kantowski considered light propagation in the `Swiss-Cheese' cosmological
model \cite{KANTO}.
Owing to the lack of exact solutions of the equations of General
Relativity that can describe an inhomogeneous universe, it was assumed that it can be described on average as a FLRW 
universe whose metric is locally perturbed by matter inhomogeneities (
although as discussed in \cite{EHLERS} how this works is not obvious at all, 
due to the non-linearity of the Einstein equations). For the limiting case, 
where a light bundle propagates far away from all matter inhomogeneities 
C.C. Dyer and R.C. Roeder obtained a modified redshift-distance relation
\cite{EHLERS}.
The exact inhomogeneous solution obtained in this paper for the metric 
and the matter energy density may be used for 
realistic investigations of light propagation in inhomogeneous space-times.

Another interesting application of the exact solution derived in this paper 
is to address the problem of Galaxy formation. Lemaitre \cite{LEMAITRE} 
predicted that an inhomogeneous Universe, with a non-zero Cosmological 
Constant, should 
contain collapsing regions spread over an expanding background (FLWR). In 
his paper he suggested that the expansion of the outer regions were caused 
by the cosmological constant repulsion while ``local'' instabilities could 
provide the mechanism for the formation of galaxies \cite{LEMAITRE}. 

The above physical applications of the exact solution to the inhomogeneous 
cosmology with the cosmological constant will be the subject of a separate 
future publication.

\section{Homogeneous Cosmology}

\subsection{Recovery of the Friedmann Equation}

In this section we will show how the Friedmann equation (FLRW limit) can be
recovered from the Szekeres-Szafron metric by judicious choice of the
various functions, parameters and constants (See Krasinski \cite{KRASINSKI}).

The starting point is the metric Eq.(\ref{metriki}),
where $e^{\beta (x,y,z,t)}$ ,$e^{\alpha (x,y,z,t)}$ are given by 
Eqns.(\ref{ebeta}),(\ref{ealpha}) respectively and $e^{-\nu}$ by:

\begin{equation}
e^{-\nu } =A(z)(x^{2}+y^{2})+2B_{1}(z)x+2B_{2}(z)y+C(z).
\end{equation}%

Here $A(z)$,$B_{1}(z)$,$B_{2}(z)$,$C(z)$ and $h(z)$ are arbitrary functions
while $K(z)$ is determined by:

\begin{equation}
AC-B_{1}^{2}-B_{2}^{2}=\frac{1}{4}\left[ h^{-2}(z)+K(z)\right] .
\end{equation}

Now by choosing:

\begin{equation}
\Phi (t,z)=f(z)R(t),\quad K(z)=K_{0}f^{2}(z),\ M(z)=M_{0}z^{3},
\end{equation}
the FLRW limit \cite{KRASINSKI} in the Goode \& Weinwright form \cite{GOODE}%
, can be derived, namely:

\begin{equation}
ds^{2}=dt^{2}-R^{2}(t)[e^{2\nu }(dx^{2}+dy^{2})+W^{2}f^{2}\nu
_{,z}^{2}dz^{2}],
\end{equation}
here,
\begin{eqnarray}
W^{2} &=&(\epsilon -Kf^{2})^{-1}, \\
e^{\nu } &=&f(z)[a(z)(x^{2}+y^{2})+2b(z)x+2c(z)y+d(z)]^{-1},
\end{eqnarray}
$\epsilon ,K$ are arbitrary constants and $R(t),f(z),a(z),b(z),c(z),d(z)$
are arbitrary functions subject to the constraint $ad-b^{2}-c^{2}=\epsilon
/4.$

If in\ addition\ we choose, $B_{1}=B_{2}=0,C=1$, $A=\frac{1}{4},$ and $%
f(z)=z $, then the FLRW limit is recovered, leading to the standard
Friedmann equation, given by:

\begin{equation}
\dot{R}^{2}=\frac{8\pi G_{N}}{3}\rho _{matter}R^{2}+\frac{c^{2}\Lambda }{3}%
R^{2}-Kc^{2},  \label{friedmann}
\end{equation}
and the matter energy density eq.(\ref{DENSITYM}), which is a solution of
the gravitational field equations, in this limit becomes:

\begin{equation}
k\widetilde{\rho }=\frac{2M_{,z}+6M\nu _{,z}}{\Phi ^{2}(\Phi _{,z}+\Phi \nu
_{,z})}=\frac{6M_{0}}{R^{3}(t)},
\end{equation}

where \ $M_{0}$ is a Friedmann mass integral \cite{KRASGR}. In Eq.(\ref%
{friedmann}) $K$ and $\Lambda $ denote the curvature coefficient, and
Cosmological Constant, respectively.

\bigskip Now by choosing suitable parameters, namely,

\begin{eqnarray}
2M &=&\frac{8\pi G_{N}}{3}\times constant \text{,}  \notag \\
K^{\prime } &=&Kc^{2},  \notag \\
\Lambda ^{\prime } &=&\Lambda c^{2}.
\end{eqnarray}
equation (\ref{friedmann}) can be rewritten in the standard Friedmann form
(Eq.(\ref{modular})),
\begin{equation}
\dot{R}^{2}=\frac{2M}{R}+\frac{\Lambda ^{\prime }}{3}R^{2}-K^{\prime },
\label{energyb}
\end{equation}
whose solution is given by:

\begin{eqnarray}
R(t) &=&\frac{M/2}{\wp (u+\epsilon ,\tau )+K^{\prime }/12},  \notag \\
t &=&\frac{1}{c}\sqrt{\frac{3}{\Lambda }}\left[ \text{Log}\left[ \frac{%
\sigma (u+\epsilon -v_{0})}{\sigma (u+\epsilon +v_{0})}\right] +2u\ \zeta
(v_{0})\right] .
\label{exxx}
\end{eqnarray}
and the cubic invariants are given by:
\begin{equation}
g_2=\frac{1}{12} K^{\prime 2}, g_3=\frac{1}{216}K^{\prime 3}-
\frac{1}{12}\Lambda^{\prime}M^2.
\label{cubinv}
\end{equation}
and $\wp(v_0)=-\frac{K^{\prime}}{12}$.

Alternatively, the solution for the scale factor and the time in terms of
Jacobi theta functions are given by:

\begin{eqnarray}
R(t) &=&\frac{M}{2}\left\{ e_{\alpha }+\frac{1}{4\omega ^{2}}\left[ \frac{%
\theta _{1}^{\prime }\left( 0\right) }{\theta _{\alpha +1}\left( 0\right) }%
\frac{\theta _{\alpha +1}\left( \gamma \right) }{\theta _{1}\left( \gamma
\right) }\right] ^{2}+\frac{K^{\prime }}{12}\right\} ^{-1}, \\
t &=&\sqrt{\frac{3}{\Lambda }}\left[ Log\left( \frac{\theta _{1}\left(
\gamma -\gamma _{0}\right) }{\theta _{1}\left( \gamma +\gamma _{0}\right) }%
\right) +2\gamma \frac{\theta _{1}^{\prime }\left( \gamma _{0}\right) }{%
\theta _{1}\left( \gamma _{0}\right) }\right] ,
\end{eqnarray}
where $\gamma =\frac{u+\epsilon }{2\omega }$, $\gamma _{0}=\frac{v}{2\omega }
$ and $e_{\alpha }$ are the three roots of the given cubic equation.

\subsubsection{Theoretical Expression for the Hubble Constant}

The theoretical expression for Hubble's Constant is given below in terms of
Weierstra\ss , or alternatively, Jacobi theta functions.

The solution of the Friedmann equation for the scale factor is given in 
parametric form by eq.(\ref{exxx}), and the derivative of the scale factor 
with respect to time $\dot{R}:=\frac{dR}{dt}$ is given by :

\begin{equation}
\dot{R}(t) =\frac{\wp ^{\prime }(u+\epsilon )}{\wp (u+\epsilon )+\frac{%
K^{\prime }}{12}}\text{ ,}
\label{dersca}
\end{equation}
where $\wp ^{\prime }(z)=\frac{\partial \wp (z)}{\partial z}.$
One can obtain (\ref{dersca}) by directly differentiating 
eq.(\ref{exxx}) with 
respect to time. Alternatively by substituting (\ref{exxx}) in eq.(\ref{friedmann}) and using the equation for the elliptic curve: 
\begin{equation}
(\wp ^{\prime }(u+\epsilon ))^2=4\wp ^{3}(u+\epsilon )-g_{2}\;\wp
(u+\epsilon )-g_{3},
\end{equation}
one rederives (\ref{dersca}). This procedure provides a consistency check 
for our calculations.

The theoretical expression for Hubble's parameter for 
a homogeneous cosmology is given by:

\begin{equation}
H(t)\equiv \frac{\dot{R}(t)}{R(t)}=-\frac{2}{M}\wp ^{\prime }(u+\epsilon ),
\label{HUBBLEPAR}
\end{equation}

We note that Hubble's Constant is proportional to $\wp ^{\prime }(u).$
Equation (\ref{HUBBLEPAR}) is remarkable since it connects the expansion
rate of the Universe, a dynamical parameter, with the cosmological
parameters and constants of nature in a highly non-linear way.

Alternatively, Hubble's Constant can be expressed in terms of Jacobi theta
functions, as follows:\qquad \qquad 
\begin{equation}
H(t)=\frac{2}{M}\frac{1}{4\omega ^{3}}\left( \frac{\theta _{2}\left( \gamma
\right) \theta _{3}\left( \gamma \right) \theta _{4}\left( \gamma \right)
\theta _{1}^{\prime 3}\left( 0\right) }{\theta _{2}\left( 0\right) \theta
_{3}\left( 0\right) \theta _{4}\left( 0\right) \theta _{1}^{3}\left( \gamma
\right) }\right) .
\end{equation}

Also the deceleration parameter $q=-\frac{\ddot{R}R}{\dot{R}^{2}}$ is given
by the following expression:

\begin{eqnarray}
q &=&\frac{\wp ^{\prime \prime }(u+\epsilon )[\wp (u+\epsilon )+\frac{%
K^{\prime }}{12}]}{\wp ^{\prime 2}(u+\epsilon )}-1,  \notag \\
&=&\frac{[6\wp ^{2}-\frac{1}{2}g_{2}][\wp (u+\epsilon )+\frac{K^{\prime }}{12%
}]}{\wp ^{\prime 2}(u+\epsilon )}-1.  \label{DECELERATION}
\end{eqnarray}


\section{Roots of the cubic in terms of the cosmological parameters and 
regions of the fundamental period parallelogram that correspond to physical 
solutions}

In this section  we shall discuss at which
regions of the complex plane the Weierstra$\ss$ function, $\wp(z)$, and 
its derivative $\wp^{\prime}$, are real.
We restrict attention in these regions of the complex plane since 
we require that the scale factor is positive and real 
$R\geq 0)$, and that the time and Hubble's Constant
are real physical quantities. Due to the 
above necessary conditions it is important to discuss the 
relevant properties of the elliptic functions
appearing in our solutions. For further formulas and 
mathematical background the reader is referred to
the book by Abramowitz \cite{ABRAMOWITZ} and the book by 
Silverman and Tate \cite{JOSEPH}.

The determination of the cosmological quantities such as the scale factor $%
R, $ \ involves finding the roots of the cubic polynonial, $4z^{3}$ $%
-g_{2}z-g_{3}$, of the elliptic curve for various values of the
cosmological parameters $\Lambda ,$ $M$ and $K$.

The three roots of the cubic, $e_{i}$ , $i=1,2,3$ can be determined
by using the algorithm developed by Tartaglia and Gardano \cite{TARTACARDA}
and their expressions in terms of $\Lambda ,$ $M,K$ are given by :

\begin{eqnarray}
e_{1} &=&\frac{1}{12}\left\{ K^{3}-18M^{2}\Lambda +6\sqrt{-K^{3}\Lambda
M^{2}+9M^{4}\Lambda ^{2}}\right\} ^{1/3}  \notag \\
&&+\frac{1}{12}\left\{ K^{3}-18M^{2}\Lambda -6\sqrt{-K^{3}\Lambda
M^{2}+9M^{4}\Lambda ^{2}}\right\} ^{1/3},  \notag \\
e_{2} &=&\frac{\rho }{12}\left\{ K^{3}-18M^{2}\Lambda +6\sqrt{-K^{3}\Lambda
M^{2}+9M^{4}\Lambda ^{2}}\right\} ^{1/3}  \notag \\
&&+\frac{\rho ^{2}}{12}\left\{ K^{3}-18M^{2}\Lambda -6\sqrt{-K^{3}\Lambda
M^{2}+9M^{4}\Lambda ^{2}}\right\} ^{1/3},  \notag \\
e_{3} &=&\frac{\rho ^{2}}{12}\left\{ K^{3}-18M^{2}\Lambda +6\sqrt{%
-K^{3}\Lambda M^{2}+9M^{4}\Lambda ^{2}}\right\} ^{1/3}  \notag \\
&&+\frac{\rho }{12}\left\{ K^{3}-18M^{2}\Lambda -6\sqrt{-K^{3}\Lambda
M^{2}+9M^{4}\Lambda ^{2}}\right\} ^{1/3},
\end{eqnarray}

\bigskip where $\rho =e^{2\pi i/3}.$

\bigskip

In general there are three types of solution depending on whether the
discriminant is either zero, positive or negative. The solutions are
outlined below:

1.\quad $\Delta =0$, \ the roots are all real and at least two of them
coincide. \ The curve is no longer an elliptic curve, since it now has a
singular point. The solutions are not given by elliptic functions and do not
have modular properties.

\bigskip

2. \quad $\Delta <0$, the roots $e_{1}$and $e_{3}$ form a complex
conjugative pair with $e_{2}$ real. These solutions are given by elliptic
functions and have modular properties.

\bigskip

3.\quad $\Delta >0$, all the roots are real and the solution is expressed in
terms of elliptic functions which have modular properties.

\bigskip

The lattice $L$ of periods $2\omega ,2\omega ^{\prime }$ associated to the
given cubic equation (that corresponds to a choice of the parameters $%
\Lambda ,M,K$) is determined by integrating certain Abelian differentials $%
\frac{dx}{y}$ \ described below.

Consider the case $\Delta <0$. In this case two of the roots $e_{1},e_{3}$
are conjugate complex and one $e_{2}$ is real.
There is a complex conjugate pair of semi fundamental periods, $\omega $ and 
$\omega ^{\prime },$ which are given by:

\begin{equation}
\omega =\int_{e_{1}}^{\infty }\frac{dz}{\sqrt{4z^{3}-g_{2}z-g_{3}}},\quad
\omega ^{\prime }=\int_{e_{3}}^{\infty }\frac{dz}{\sqrt{4z^{3}-g_{2}z-g_{3}}}%
.
\end{equation}

For $\Delta >0$, all the three roots $e_{1,}e_{2},e_{3}$ of $%
4z^{3}-g_{2}z-g_{3},$ are real and if the $e_{i}$ are ordered so that $%
e_{1}>e_{2}>e_{3}$ we can choose the periods as:

\begin{equation}
\omega =\int_{e_{1}}^{\infty }\frac{dt}{\sqrt{4t^{3}-g_{2}t-g_{3}}},\quad
\omega ^{\prime }=i\int_{-\infty }^{e_{3}}\frac{dt}{\sqrt{%
-4t^{3}+g_{2}t+g_{3}}}.
\end{equation}
In this case $\omega $ and $\omega ^{\prime }$ are real and totally
imaginary respectively. The period ratio $\tau $ is defined by $\tau =\omega
^{\prime }/\omega .$

The three roots of the cubic can be expressed in terms of the Jacobi 
theta functions \cite{ABRAMOWITZ},
and therefore have non-trivial transformation properties under the 
modular group $SL(2,Z)$. For instance under the modular transformation
$\tau\rightarrow -\frac{1}{\tau}$, $e_1 \rightarrow e_3$, $e_3\rightarrow 
e_1$ and $e_2$ is invariant. The roots are modular invariant under subgroups 
of $SL(2,Z)$. In fact the factor group $PSL(2,Z)/\Gamma_2$ where $\Gamma_2$ 
is the modular group of level 2 is isomorphic to $S_3$, the permutation 
group with six elements acting on the letters $e_1,e_2,e_3$. 
One can construct modular invariant quantities using the above transformation 
properties. The combination of cosmological parameters given by 
the relation $(e_2-e_3)(e_1-e_3)^{-1}$ is the absolute invariant of 
$\Gamma_2$. 

\subsubsection{The Fundamental Period Parallelogram for $\Delta <0$}

\begin{figure}
\epsfxsize=6.5in
\epsfysize=7.5in
\epsffile{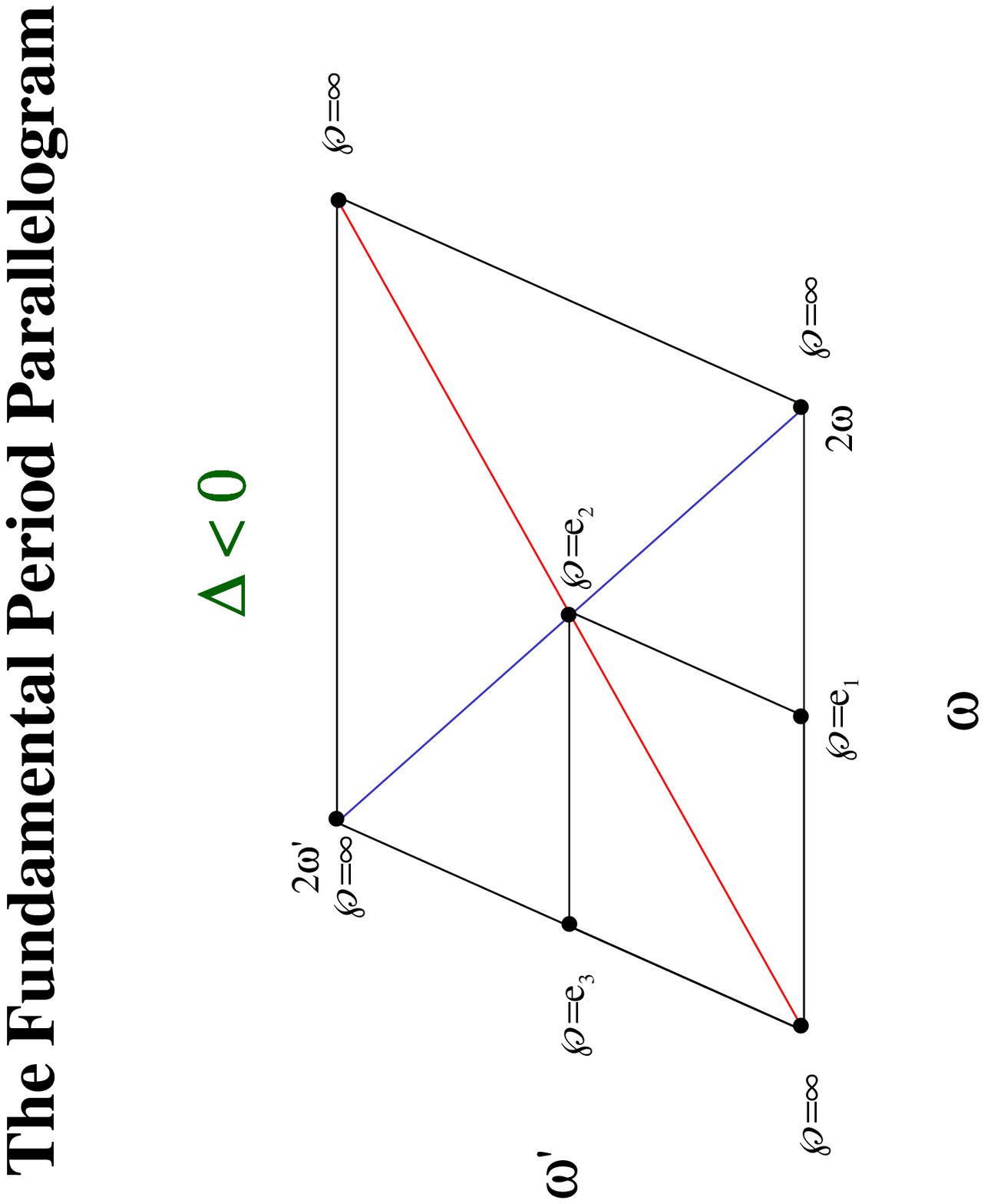}
\caption{Fundamental domain for $\Delta<0$}
\label{FPPNEGD}
\end{figure}

The Fundamental Period Parallelogram (FPP) for a typical solution, $\Delta
<0,$ is shown in figure \ref{FPPNEGD}. The figure also shows the position of
the real root $e_{2}$, and the two complex conjugatives $e_{1}$ and $e_{3}$.
In order to obtain physically meaningful (i.e. real) values for the scale
factor $R$, we restrict ourselves to a set of solutions for the 
Weierstra\ss function and its derivative that lie along the diagonals
of the FPP.
Mathematically this can be described by:\newline

The Weierstra$\ss$ function $\wp \left( z\right) =\wp \left( u+\epsilon
\right) ,$ $z=u+\epsilon $, where $\epsilon $ is a constant of integration,%
\newline
is real when the argument $z=\rm{Re}\left( z\right) +i\rm{Im}\left(
z\right) $ $=m\left( \omega +\omega ^{\prime }\right) +n\left( \omega
-\omega ^{\prime }\right) ,$ with$,m,n\in $ $R.$\newline

Note that the Weierstra\ss\ function and its derivative are real for the
diagonal that goes left to right, and real and totally imaginary,
respectively, for the right to left diagonal. \ As $z$ varies along
diagonals of period parallelograms from $0$ to $2(\omega +\omega ^{\prime })$
$\wp (z)$ decreases from $+\infty $ to $e_{2}$ to $\infty .$

We also note that the zeros of the derivative of the Weierstra\ss\ function $%
\wp ^{\prime }(x)=2\sqrt{(\wp -e_{1})(\wp -e_{2})(\wp -e_{3})}$ are at the
point $\omega ,\omega ^{\prime },\omega +\omega ^{\prime }$ of the FPP in
the complex plane. From the Euclidean uniformization 
$f:u\rightarrow (\wp(u),\wp^{\prime}(u)),$ u a complex number,
 of the Weierstra\ss\ elliptic curve these points
correspond to the three points of the elliptic curve of order two \footnote{%
A point $P$ is said to have \emph{order m} if $mP=\underset{m\text{\ summands%
}}{\underbrace{P+P\cdots +P}}$ $=\mathcal{O}$ but $m^{\prime }P\not=\mathcal{%
O}$ for all integers 1$\leq m^{\prime }<m,$ and $\mathcal{O}$ is the
identity element of the elliptic curve group. If such an $m$ exists, then $P$
has \emph{finite order }; otherwise it has \emph{infinite order. }In order
to find points of order dividing $m,$ we look for points in the FPP such
that $mP\in L.$ (Of course, we send the points in $L$, which are the poles of 
$\wp$, to $\mathcal{O}$).}.

\bigskip

\subsubsection{\protect\bigskip The Fundamental Period Parallelogram for $%
\Delta >0$}

\begin{figure}
\epsfxsize=6.5in
\epsfysize=7.5in
\epsffile{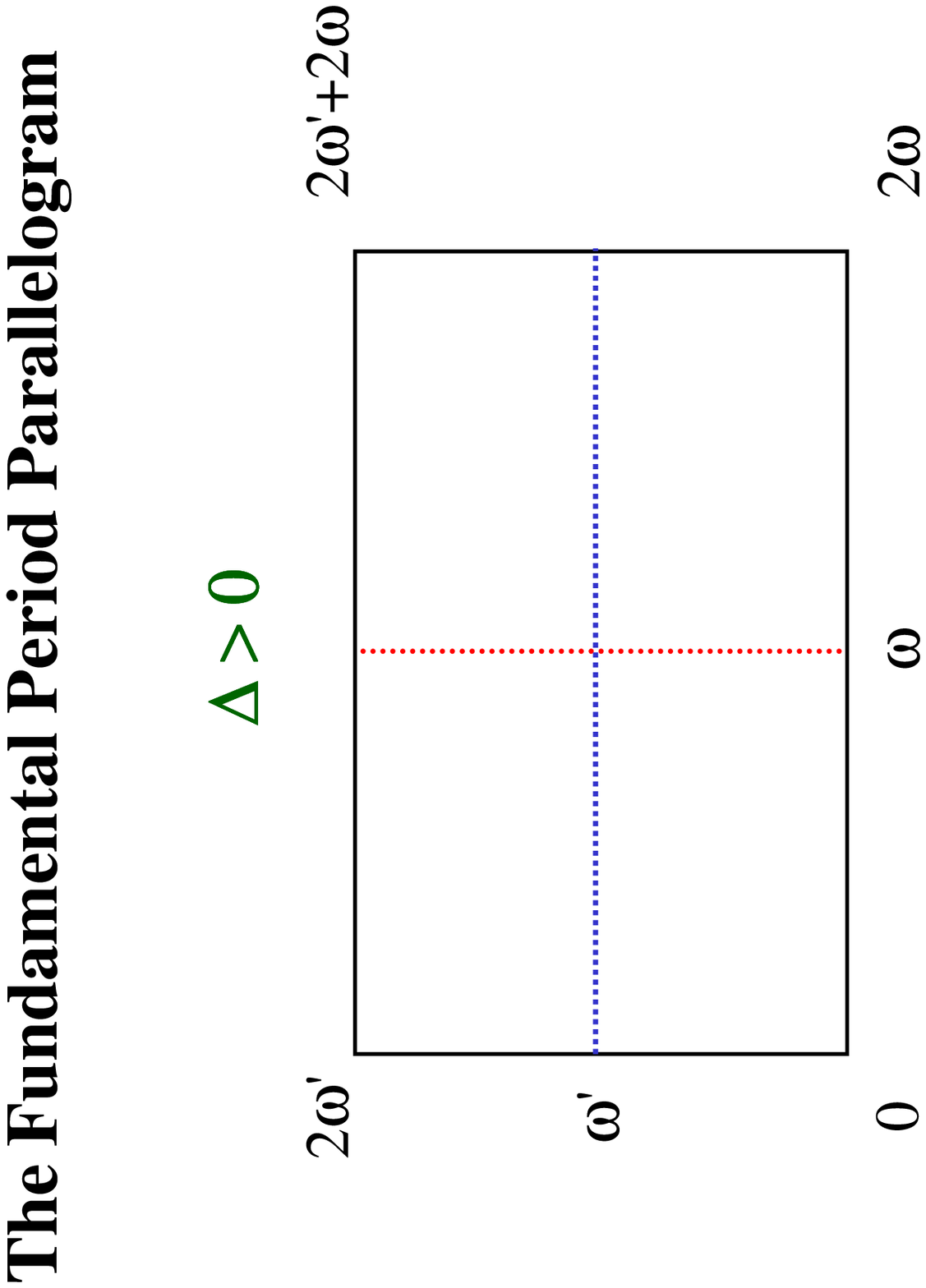}
\caption{Fundamental domain for $\Delta>0$}
\label{FPPPOSID}
\end{figure}

The Fundamental Period Parallelogram (FPP) for, $\Delta >0,$ is shown in
figure (\ref{FPPPOSID}). There are three real roots $e_{1},e_{2},e_{3}\in R$%
, and $\omega ,\omega ^{\prime }$ are real and totally imaginary,
respectively. The real values of the Weierstra\ss\ function lie on the
vertical and horizonal lines defined by:\newline

$z=u+\epsilon $, $\wp \left( z=a+ib\right) \in R$\newline

for $a=0$ or $\omega $ and $b=0$ or $\omega ^{\prime }.$\newline

\bigskip

The Weierstra\ss\ derivative is real on the horizontal lines and totally
imaginary on the vertical ones. More specifically, in the interesting case
of $g_{2}=\frac{K^{\prime 2}}{12}\not=0,g_{3}=0,$ the resulting lattice is a
square lattice with $\omega ^{\prime }=i\omega ,$ i.e., the lattice $L$ is
the Gaussian integer lattice expanded by a factor of $\omega ^{\prime }.$ In
this case we will see in section (4.1.2) that as $z$ travels along the
straight line from $\omega ^{\prime }$ to $\omega ^{\prime }+2\omega ,$ the
point ($x,y)=(\wp (z),\wp^{\prime} (z))$ moves around the real points of
the elliptic curve $y^{2}=4x^{3}-g_{2}x$ between $-K^{\prime }/(4\sqrt{3})$
and $0,$ \ and as $z$ travels along the straight line from $0$ to $2\omega ,$
the point $(x,y)=(\wp (z),\wp^{\prime} (z))$ travels through all the real
points of this elliptic curve which are to the right of $(\frac{K^{\prime }}{%
4\sqrt{3}},0)$.

We also note that in the case $\Delta \not=0,$ the points of order 3
correspond to the inflection points of the elliptic curve. We found that at
these points and for a positive Cosmological Constant the deceleration
parameter tends to $-1$, leading to the asymptotic value for the Hubble's
Constant given by $\Lambda c^{2}=3H_{\infty }^{2}$. Geometrically, points of
order three are the points where the tangent line to the cubic has a triple
order contact.

\bigskip

\newpage

\section{Analysis and General Results}

\bigskip

\subsection{\protect\bigskip Introduction}

Before presenting the results for the scale factor $R(t)$ in the general
case, in which both the discriminant $\Delta $ and the cubic invariants $%
g_{2}$ and $g_{3}$ are non-zero, we shall present three interesting cases in
which one of the above mentioned parameters vanishes. Each of these cases
has a particular physical interpretation and special mathematical properties
which, with the exception of $\Delta =0,$ have not been discussed in the
literature before. The three cases considered are:

\begin{eqnarray}
\Delta &=&0,\ \quad g_{2}\not=0,\quad g_{3}\not=0,  \label{ORDINARY} \\
\Delta &>&0,\quad \ g_{2}\not=0,\quad g_{3}=0,  \label{AVARIETY1} \\
\Delta &<&0,\quad \ g_{2}=0,\quad g_{3}\not=0.  \label{AVARIETY2}
\end{eqnarray}

The presentation and ordering of the cosmological results has been dictated
by the mathematical properties of the elliptic curve: the mathematics
determines the nature of the allowed physics.

\subsection{Analysis}

\subsubsection{\protect\bigskip The Case of Vanishing Discriminant, $\Delta
=0$}

There are three limiting cases in which the discriminant $\Delta $ of the
cubic equation vanishes. This occurs when at least two of the roots $e_{i}$
coincide. As one can see from the form of the cubic invariants, $g_{2},$ $%
g_{3}$ Eq.(\ref{cubinv}), the discriminant $\Delta $ vanishes when
either (or both) of $\Lambda ,M$ is zero.

In the third case, $\Delta $ is identically zero for the choice of
parameters given by:

\begin{equation}
\Lambda 
{\acute{}}%
=\frac{1}{9}\frac{K%
{\acute{}}%
^{3}}{M^{%
{\acute{}}%
2}}\equiv \Lambda _{Critical}^{%
{\acute{}}%
}.  \label{CRITICAL}
\end{equation}%
In all three cases the solutions for the cubic equation are no longer
described by elliptic functions and they lose their modular properties (the
Weierstra\ss\ functions degenerate). The resulting solutions are expressed
in terms of elementary functions and represent the standard textbook
solutions of Cosmology \cite{OHANIAN}.

\subsubsection{\protect\bigskip Class of Solutions with a non-zero
Discriminant and One of the two Cubic Invariants zero.}

This section examines the two special cases where one of the cubic
invariants $g_{2},g_{3}$ is zero for a non-zero discriminant. It will be
shown that the resulting solutions are associated with special elliptic curves and have important
mathematical and physical properties.

The first case is when $g_{2}=0$ , i.e. $K%
{\acute{}}%
=0$ (case of a flat universe)$,$ $g_{3}=-\frac{1}{12}\Lambda ^{\prime
}M^{2}\not=0$ . Here the discriminant $\Delta $ is negative and the
resulting two types of solutions are characterised by the sign of $\Lambda ,$
which in mathematical terms correspond to the fixed points in the upper half
complex plane ($\tau =\pm \frac{1}{2}+\frac{\sqrt{3}}{2}i).$

The second case is when $g_{3}=0,$ i.e. $\Lambda 
{\acute{}}%
=\frac{1}{18}\frac{K%
{\acute{}}%
^{3}}{M^{2}}$ \ and $g_{2}$ $\not=0.$ Here $\Delta =g_{2}^{3}>0$ and the
resulting solutions are characterised by the sign of the curvature constant $%
K%
{\acute{}}%
.$ The solutions correspond to the other fixed point in the upper half
complex plane $\left( \tau =i\right) .$ These special points\ for\ $\tau $
in the upper half complex plane have significant symmetry properties which
will be discussed in later sections of this paper.

\bigskip

\paragraph{\protect\bigskip Euclidean Universe \ $(\Delta <0,\ g_{2}=0,\
g_{3}\not=0)$\\}

\bigskip

For $K=0,\Lambda \not=0,M\not=0,$ $g_{2}$ vanishes while $g_{3}$ and the
discriminant $\Delta $ are both non-zero. This is the case of the Euclidean
Universe where the solutions are described by elliptic functions.
In  this case:
\begin{equation}
g_{2}=0,\text{ \ }g_{3}=\frac{-1}{12}\Lambda ^{\prime }M^{2},\text{ \ }%
\Delta =-27g_{3}^{2}<0,
\end{equation}
the roots are given by:

\begin{eqnarray}
e_{1} &=&\frac{1}{12}\sqrt[3]{-36\Lambda ^{\prime }M^{2}},  \notag \\
e_{2} &=&\frac{\rho }{12}\sqrt[3]{-36\Lambda ^{\prime }M^{2}},  \notag \\
e_{3} &=&\frac{\rho ^{2}}{12}\sqrt[3]{-36\Lambda ^{\prime }M^{2}},
\end{eqnarray}
and the cubic equation becomes:

\begin{equation}
\wp 
{\acute{}}%
^{2}(z)=4\wp ^{3}\left( z\right) -g_{3}.  \label{diofantis}
\end{equation}

For this case the discriminant is always negative and that there are two
types of solutions depending upon the sign of the Cosmological Constant.
Interesting physics and mathematics arise when equation (\ref{diofantis}) is
solved at the zeros \footnote{%
A beautiful formula for determining the zeros of the Weierstra\ss\ function $\wp $ has been obtained by Eichler and Zagier \cite{DON} and is 
given by: $\wp (\tau ,z)=0\Leftrightarrow
z=\lambda \tau +\mu \pm \left( \frac{1}{2}+\frac{\log (5+2\sqrt{6})}{2\pi i}%
+144\pi i\sqrt{6}\int\limits_{\tau }^{i\infty }(t-\tau )\frac{\Delta (t)}{%
E_{6}(t)^{3/2}}dt\right) $} of the Weierstra\ss\ function, $\wp (u_{0})=0$,
it reduces to:

\begin{equation}
\wp ^{\prime 2}(u_{0})=-g_{3}.
\end{equation}

\bigskip For $\Lambda ^{\prime }>0$ the defining equation becomes:

\begin{eqnarray}
\wp ^{\prime }(u_{0}) &=&i\sqrt{g_{3}}  \notag \\
\wp ^{\prime }(u_{0}) &=&i\sqrt{-\frac{\Lambda ^{\prime }M^{2}}{12}}=-\frac{M%
}{2}\sqrt{\frac{\Lambda ^{\prime }}{3}},
\end{eqnarray}

leading to the expression for Hubble's Constant given by:

\begin{eqnarray}
H &=&-\frac{2}{M}\wp ^{\prime }(u_{0}),  \notag \\
&=&\sqrt{\frac{\Lambda ^{\prime }}{3}},  \notag \\
&=&c\sqrt{\frac{\Lambda }{3}}.
\end{eqnarray}%
This is the expression for the Hubble parameter in de-Sitter phase \textbf{\ (%
}inflationary scenario\textbf{). }\ At the point of the complex plane that
corresponds to zero of the Jacobi modular
form \textbf{\ }$\wp ,$ a point of order 3, the flat universe scenario 
\textbf{\ \footnote{%
In fact, as we shall see shortly, \ the asymptotic value $H_{\infty }$ of
the Hubble parameter for $\Lambda >0$ is reached also in models with
non-zero $K.$} }predicts the ultimate de-Sitter phase, given by:

\begin{equation}
\Lambda c^{2}=3H_{\infty }^{2},  \label{Asymptotic}
\end{equation}
where $H_{\infty }$\textbf{\ }is the limiting value of the Hubble's Constant
as \textbf{\ }$t->\infty ,$\textbf{\ }and \textbf{\ }$R->\infty .$ Equation (%
\ref{Asymptotic}) connects the Cosmological Constant $\Lambda $, with the
asymptotic expansion rate, itself a dynamical parameter.

\begin{figure}
\epsfxsize=6.5in
\epsfysize=7.5in
\epsffile{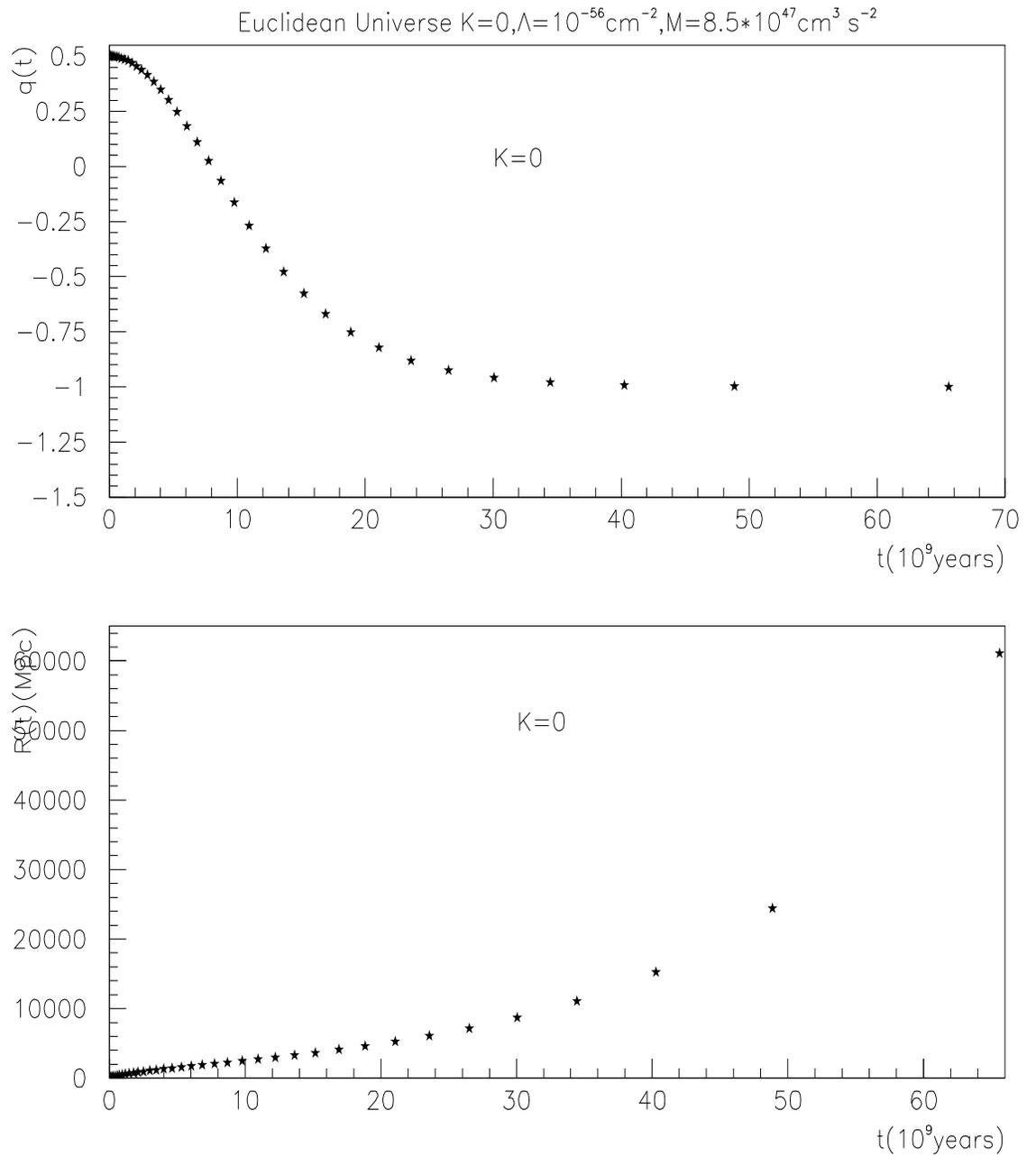}
\caption{Deceleration parameter
and scale factor in the Euclidean Universe, with parameters $\Lambda
=10^{-56}cm^{-2},M=8.5\times 10^{47}cm^{3}s^{-2},K=0.$}
\label{DECEL}
\end{figure}

As emphasised earlier, the limiting value (\ref{Asymptotic}) is achieved at
the 3-division points of the elliptic curve that correspond to zeros of the $%
\wp $ function, while the deceleration parameter $q$ given by Eq.(\ref%
{DECELERATION}) in this limit tends to $-1.$ The deceleration parameter is
shown in fig.(\ref{DECEL}). There $q$ takes values of $0.5$ when
the matter density is maximum, zero at the inflection points and $-1$ at
the points where the Cosmological Constant dominates the dynamics of the
universe.



For $\Lambda ^{\prime }<0$ the defining equation becomes:

\begin{eqnarray}
\wp ^{\prime }(u_{0}) &=&\sqrt{g_{3}}, \\
\wp ^{\prime }(u_{0}) &=&i\sqrt{\frac{\Lambda ^{\prime }M^{2}}{12}}=\frac{M}{%
2}\sqrt{\frac{\Lambda ^{\prime }}{3}},
\end{eqnarray}
leading to a Hubble's Constant given by:

\begin{equation}
H=ic\sqrt{\frac{\Lambda }{3}}.
\end{equation}

In the Euclidean Universe scenario the lattice $L$ is highly symmetrical in
that it admits the property of \emph{complex multiplication (}by $\tau
=e^{2\pi i/3})$\textbf{\ }and the Weierstra\ss\ function $\wp $ obeys the
functional relationship:

\begin{equation}
\wp \left( \tau u\right) =\tau ^{-2}\wp \left( u\right) .
\end{equation}

The above relationship has been verified, numerically,. The associated
elliptic curve has absolute modular invariant $j=0\footnote{%
The absolute modular invariant is defined as: $j\equiv 1728\times \frac{%
g_{2}^{3}}{\Delta }=1728\times \frac{g_{2}^{3}}{(g_{2}^{3}-27g_{3}^{2})}.$
We discuss in more detail its properties in section 7.}$.

\bigskip

The scale factor evolution and other cosmological parameters in the special
case of Euclidean Universe will be discussed in more detail below.

An interesting scenario, which is also favoured by recent observational data
(Type Ia Supernovae, Cosmic Microwave background radiation experiments) is
the case with $\Lambda >0.$ In the second graph of fig.(\ref{DECEL}) we have
plotted the scale factor versus time for the choice of cosmological
parameters: $\Lambda =10^{-56}$cm$^{-2},M=8.5\times 10^{47}cm^{3}s^{-2},K=0.$

\bigskip

\bigskip

\bigskip

\begin{figure}
\epsfxsize=6.5in
\epsfysize=7.5in
\epsffile{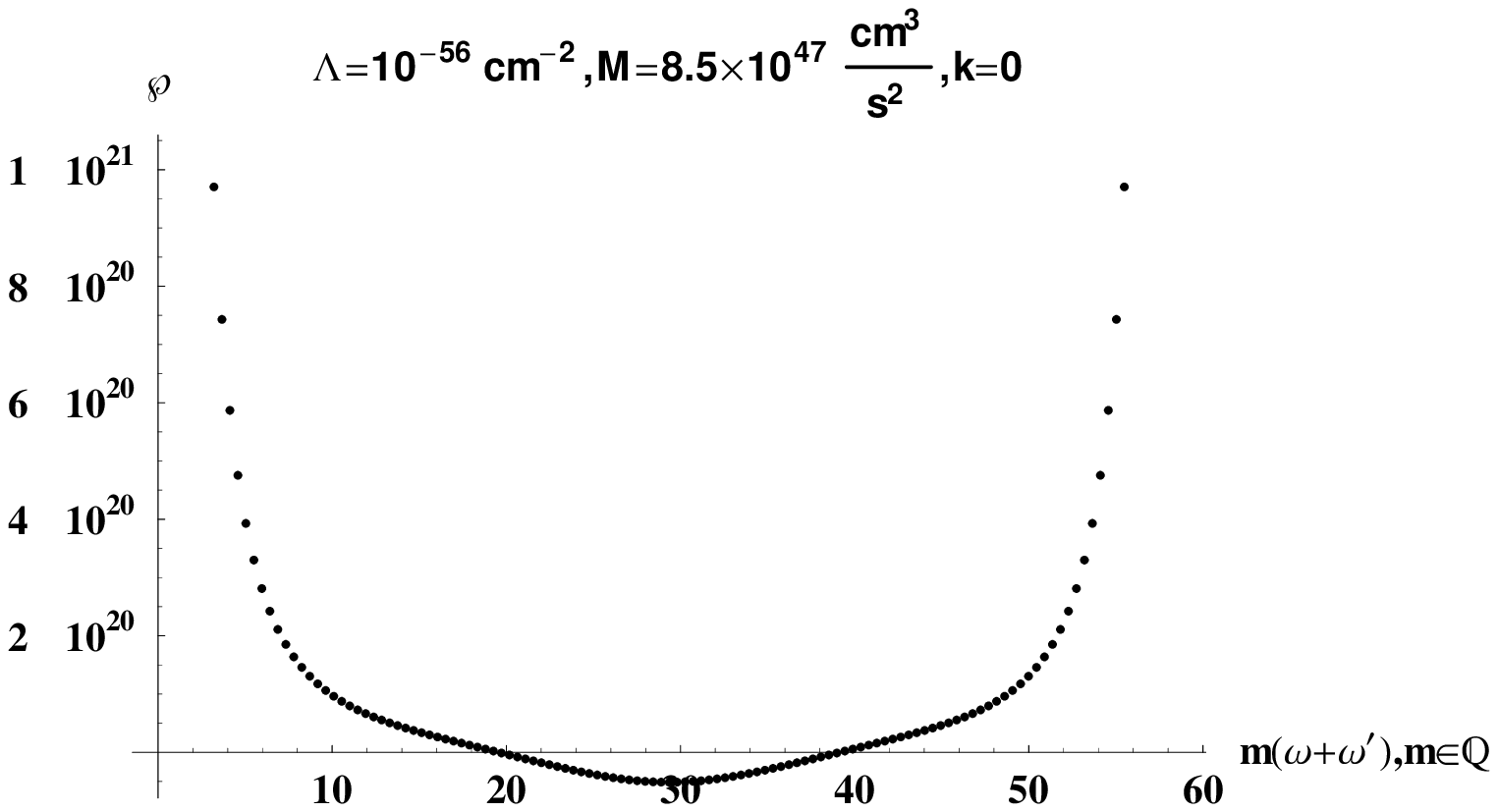}
\caption{Values of the
Weierstra \ss\ $\wp $function along the diagonal ($\protect\omega +\protect%
\omega ^{\prime })$ of the FPP.}
\label{WEIRFLATLP}
\end{figure}

Here we have integrated (\ref{energyb}) along the diagonal $m(\omega +\omega
^{\prime }),m\in R$ of the FPP in the complex plane. Since $\Delta <0$ there
is one real (negative) root $e_{2}$ and two complex conjugate roots. At the
2-division point ($\omega +\omega ^{\prime }),$ $\wp (\omega +\omega
^{\prime })=e_{2}.$ Fig.(\ref{WEIRFLATLP}) shows the behaviour of the
Weierstra\ss\ $\wp $ function along the diagonal of the FPP. As we can see
the $\wp (u)$ function starts from $\infty ,$ passes through a zero, and
reaches the negative real root $e_{2}$ as $u$ varies from $0,$ through $%
v_{0} $ and onto $\omega +\omega ^{\prime }.$ When $u\longrightarrow v_{0},$ 
$R$($t)\longrightarrow \infty ,$ $t\longrightarrow \infty $, Hubble's
Constant goes to the limiting value given by, $H_{\infty
}^{2}\longrightarrow c^{2}\frac{\Lambda }{3},$ this is the inflationary
scenario.

\bigskip

\bigskip

\bigskip

\bigskip

\begin{figure}
\epsfxsize=6.5in
\epsfysize=7.5in
\epsffile{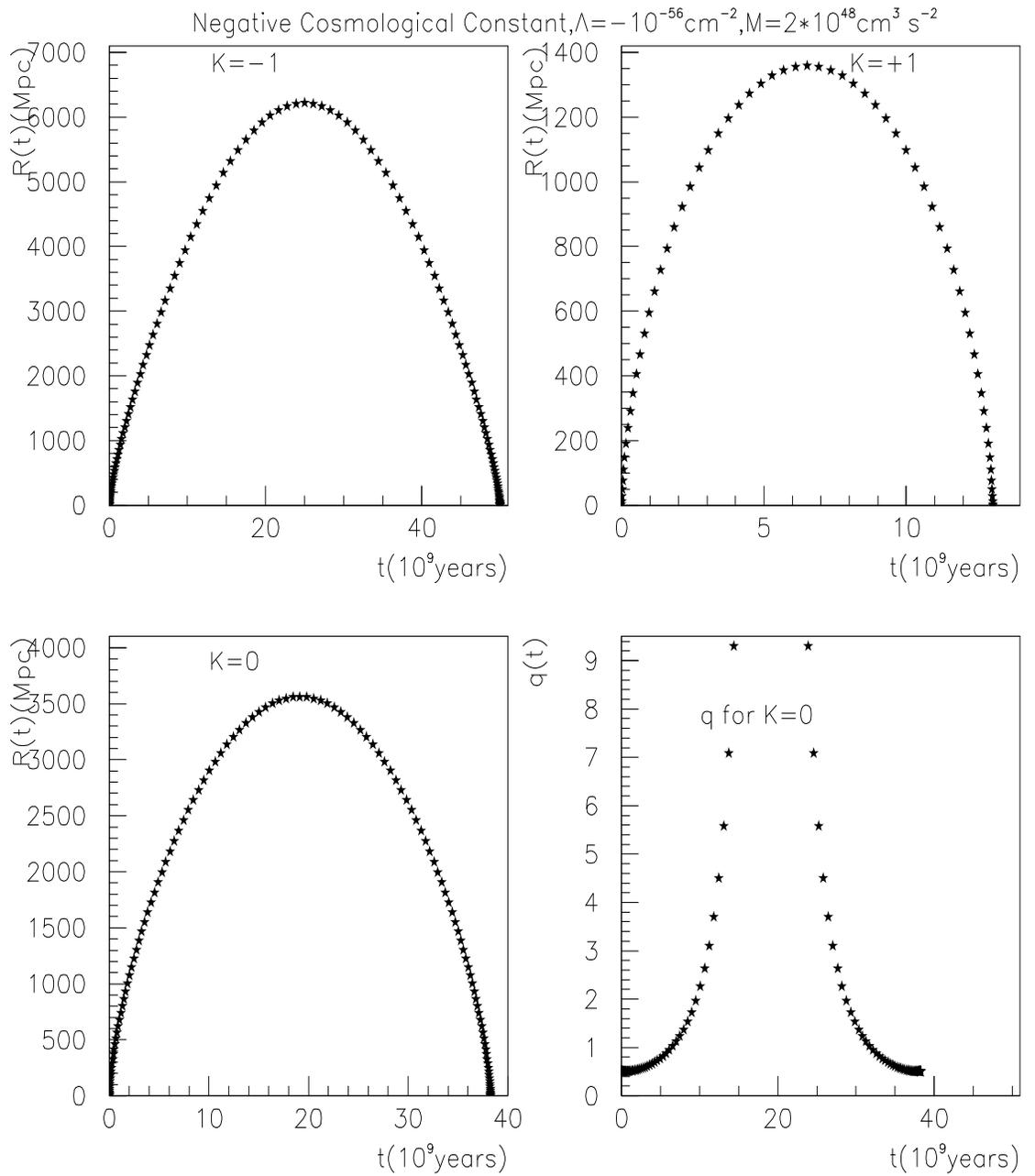}
\caption{Periodic solutions for $\Lambda<0$.}
\label{NEGATIVELAMBDA}
\end{figure}

\bigskip Finally we note that when the Cosmological Constant is negative we
obtain an oscillatory solution see fig.(\ref{NEGATIVELAMBDA}). The scale
factor has a maximum value at the 2-division point $\omega +\omega ^{\prime
} $ in the FPP. \ At this point the $\wp $ function is equal to the real
root , i.e. $\wp (\omega +\omega ^{\prime })=e_{2},$ and the first
derivative $\wp ^{\prime }$ vanishes leading to a zero value for the Hubble
Constant as expected.

\paragraph{Non-Euclidean Universe $\ (\Delta >0,g_{2}\not=0,g_{3}=0)$}

The second special case is given by the following conditions:

\begin{equation}
g_{2}=\frac{K^{\prime ^{2}}}{12},\text{ \ }g_{3}=0,\text{ \ }\Delta
=g_{2}^{3}.
\end{equation}

In this case $\Delta >0$ and there are three real roots, one of which is
zero and the other two are are equal in magnitude and opposite in sign. The
elliptic curve is given by the expression:

\begin{equation}
y^{2}=4x^{3}-g_{2}x,
\end{equation}

or in the Weierstra\ss\ parametrisation:

\begin{equation}
\wp 
{\acute{}}%
^{2}(z)=4\wp ^{3}\left( z\right) -g_{2}\wp \left( z\right) .
\label{gaussian}
\end{equation}

The vanishing of $g_{3}$ leads to the expression for the cosmological
parameters, given by:

\begin{equation}
\Lambda ^{\prime }=\frac{1}{18}\frac{K%
{\acute{}}%
^{3}}{M^{2}}=\frac{1}{2}\Lambda _{crit}^{\prime }
\end{equation}

The sign of the Cosmological Constant depends on the sign of the curvature
coefficient. Also, the elliptic curve has the property of complex
multiplication (CM), in that the square lattice $L$ admits complex
multiplication by $i.$ The Weierstra\ss\ function $\wp (x)$ satisfies the
functional relationship:

\begin{equation}
\wp (ix)=i^{-2}\wp(x)=-\wp (x).
\end{equation}

The absolute modular invariant function for the elliptic curve (\ref%
{gaussian}) has the value $j=1728
$, and since the elliptic
curve has the property of CM the absolute invariant is an algebraic integer.

There is a very interesting physical feature: at the point $v_{0}$ that
inverts the equation $\wp (v_{0})=-\frac{K^{\prime }}{12},$ the derivative
of the Weierstra\ss\ function satisfies the elliptic equation, namely:

\begin{equation}
\wp ^{\prime 2}(v_{0})=4\times \left( -\frac{K^{\prime }}{12}\right) ^{3}-%
\frac{K^{\prime 2}}{12}\left( -\frac{K^{\prime }}{12}\right) =\frac{%
8K^{\prime 3}}{12^{3}},
\end{equation}

or

\begin{equation}
\wp ^{\prime 2}(v_{0})=\frac{\Lambda ^{\prime }M^{2}}{12},
\end{equation}

at this point the Hubble's Constant becomes:

\begin{equation}
H=-\left( \frac{2}{M}\right) \wp ^{\prime }(v_{0})=-\sqrt{\frac{\Lambda
^{\prime }}{3}}=-c\sqrt{\frac{\Lambda }{3}}.
\end{equation}

Therefore we arrive at the de-Sitter functional relationship for the
Hubble's constant for $\Lambda >0$, and again the limiting value, $H_{\infty
}$, \ for the Hubble's Constant is recovered Eq.(\ref{Asymptotic}).

\begin{figure}
\epsfxsize=6.5in
\epsfysize=7.5in
\epsffile{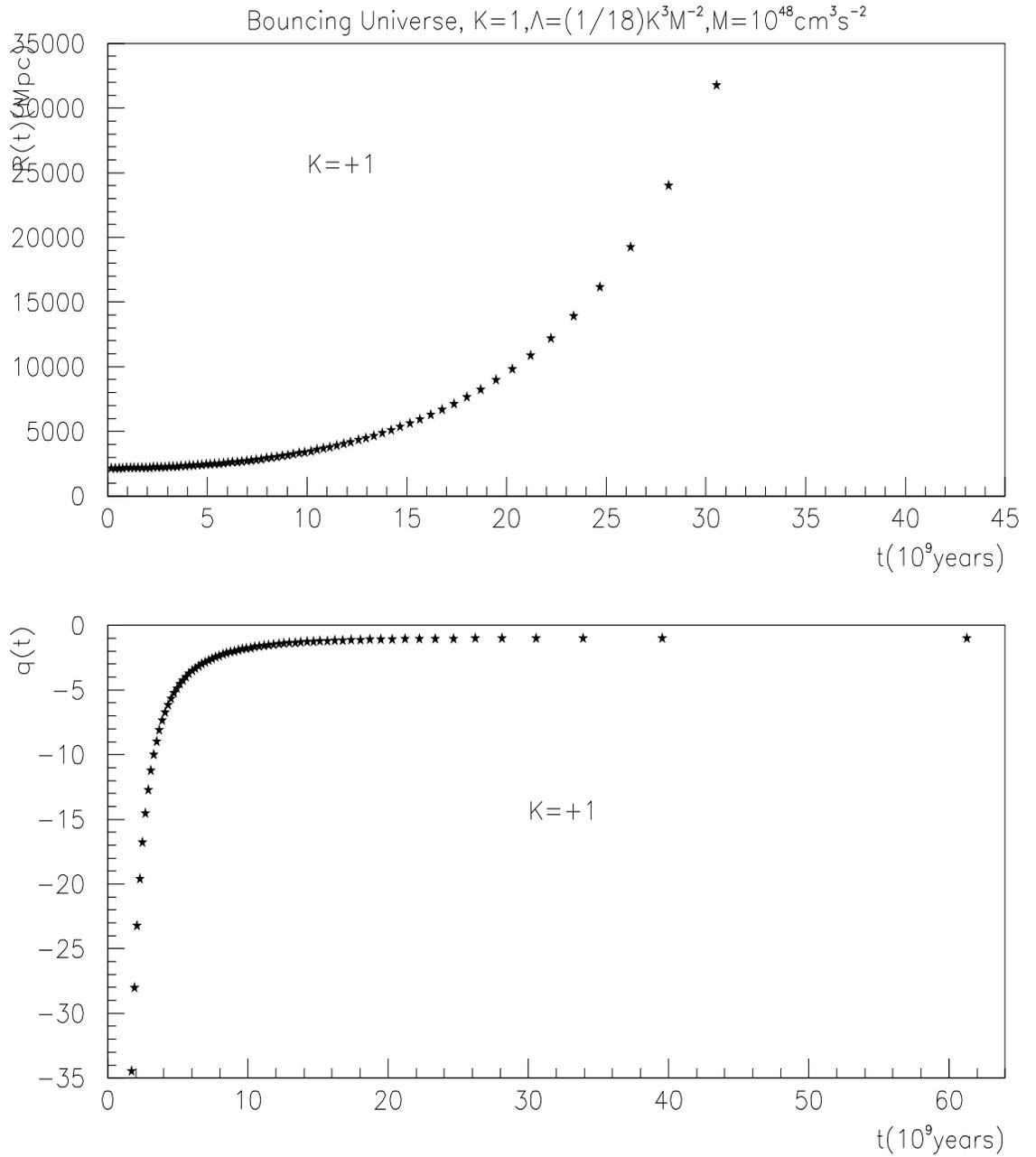}
\caption{Scale factor and
deceleration parameter versus time in the bouncing universe for 
$K=+1,M=10^{48}cm^{3}s^{-2},\Lambda =\frac{1}{18}K^{\prime 3}M^{2},$ in the
special case $g_{3}=0$.}
\label{BOUNCE}
\end{figure}

In fig.(\ref{BOUNCE}) we have plotted the scale factor versus time for the
choice of parameters: $K=1,$ $M=3.5\times 10^{48}cm^{3}s^{-2},$ $\Lambda =%
\frac{K^{\prime 3}}{18M^{2}}.$ Here we have integrated the Friedmann
equation in the region, ($\omega +\omega ^{\prime }$$\rightarrow$ $\omega
^{\prime }+2\ \omega )$ of the FPP in the complex plane, where, $\wp (u)$
lies between $e_{3}=-\frac{K^{\prime }}{4\sqrt{3}}$ and $e_{2}=0.$ In this
``bouncing'' universe scenario, the scale factor starts at a minimum value, $%
R_{\min }(t),$ and then expands exponentially forever. Interestingly, such
models do $\emph{not}$ suffer from a $t=0$ singularity.

\bigskip

\bigskip

The deceleration parameter $q,$ versus time, is also plotted in fig (\ref%
{BOUNCE}). In this case the deceleration parameter is always negative see
second graph.

\bigskip

\bigskip

\bigskip

\begin{figure}
\epsfxsize=4.5in
\epsfysize=4.5in
\epsffile{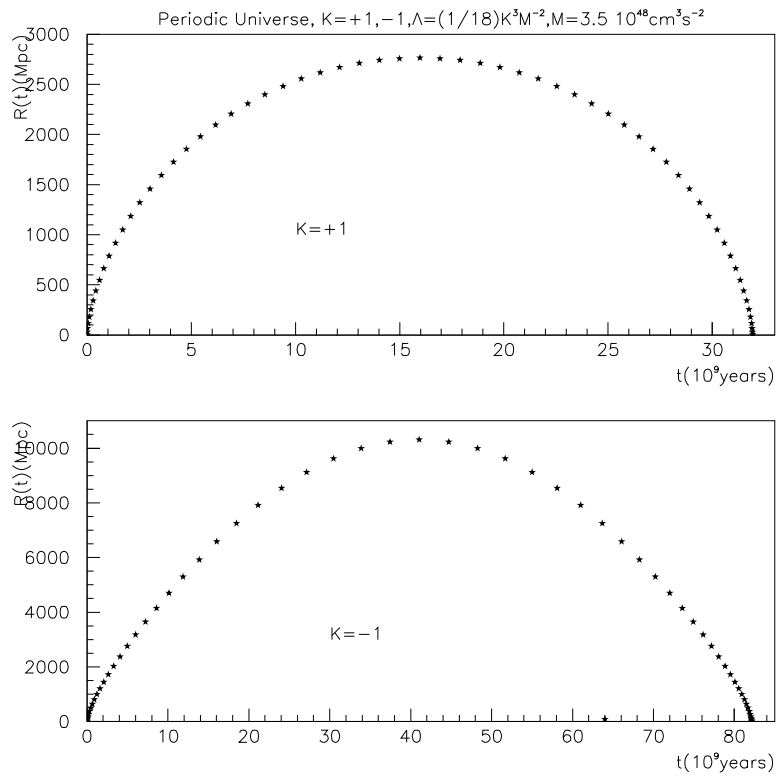}
\caption{Periodic solution for $\Lambda >0$ (first graph) 
and $\Lambda<0$ (second graph) in the special case $g_{3}=0.$}
\label{gzeroosclp}
\end{figure}

Fig.(\ref{gzeroosclp}) (first graph) presents results obtained by 
integrating eq.(\ref%
{energyb}), for the same values of the physical parameters $\Lambda ,M,K,$
in the region of the complex plane from $0$ to $2\omega .$ In this case a
periodic solution is obtained, and in this region of the FPP, the Weierstra%
\ss\ function is positive, $\wp (u)>e_{1}=\frac{K^{\prime }}{4\sqrt{3}}.$ 
For $\Lambda <0$ an oscillatory solution is  
obtained (second plot) of Fig.(\ref{gzeroosclp}).

\bigskip

\bigskip

In both cases $\Lambda =\frac{1}{2}\Lambda _{crit},$ where $\Lambda _{crit}$
is given by equation (\ref{CRITICAL}).

\subsection{General Results $(\Delta \not=0,g_{2}\not=0,g_{3}\not=0)$}

\subsubsection{\protect\bigskip\ Approach}

The results will be presented in three different classes determined by the
sign and magnitude of the Cosmological Constant. For each of the classes
arbitrary, but realistic, sets of cosmological parameters have been chosen ($%
\Lambda ,M,K$). Here the pseudomass and Cosmological Constant parameters
always have a non-zero value such that $g_{2},g_{3}$ and $\Delta $ are all
non-zero. The types of solutions can be classified as:

\bigskip \emph{Periodic or Oscillating Universe,}

\bigskip

\emph{Asymptotic Inflationary Universe, }

\bigskip

\emph{Periodic and Bouncing Universe }(for $0<$\emph{\ }$\Lambda <\Lambda
_{crit}$ and $K=1$)

\bigskip

Solving the Friedmann equation exactly leads to a set of interdependent
cosmological physical quantities:

\bigskip

Scale Factor

Age of the Universe

Hubble's Constant

Deceleration Parameter.

For each class of solution we have picked (fixed) values of the cosmological
parameters that give ''good'' results for the cosmological values.
Qualitative similar results to our own for the scale factor have been found
in Ref.\cite{STECKBRIEF}.

\bigskip\ In the following section we present our results for the scale
factor, Hubble's Constant and deceleration parameter.

\subsubsection{\protect\bigskip Periodic or Oscillating Universe}

For a negative Cosmological Constant the solutions are always periodic. We
have chosen a representative set of cosmological parameters given by:

\begin{equation}
K=0,\pm 1,\quad M=2.0\ 10^{48}cm^{3}s^{-2},\quad \Lambda =-10^{-56}cm^{-2}.
\label{PHOENIX}
\end{equation}

The resulting solutions are shown in fig(\ref{NEGATIVELAMBDA}).
We have repeated the calculations for a range of values of the cosmological 
constant ($-10^{-52}{\rm cm^{-2}}\leq \Lambda \leq 10^{-52} {\rm cm^{-2}}$)
and a range of values for the parameter $M$. The values of the parameters 
in (\ref{PHOENIX}) and in following sections were chosen in order that
the resulting cosmological solutions were as realistic as possible (i.e. 
not too small scale factor, or age of the Universe). 

The solutions show a general trend of increasing scale factor and age of the
universe as $K$ goes through 1, 0 and -1. As expected, shown in fig.(\ref%
{NEGATIVELAMBDA}), the deceleration parameter is always positive. For this
choice of the cosmological parameters, a ''good'' set of results, for $K=0$,
is given by:

\bigskip

\begin{tabular}{ll}
Scale Factor & 2,901 Mpc \\ 
Age of Universe & 10.01 Billion Years \\ 
Hubbles Constant & 49.4 $Kms^{-1}Mpc^{-1}$ \\ 
Deceleration Parameter & 2.26%
\end{tabular}

\subsubsection{Asymptotic Inflationary Universe}

\bigskip This scenario used the following set of Cosmological Parameters:

\begin{equation}
K=0,\pm 1,\ M=8.5\times 10^{47}-4\times 10^{49}cm^{2}s^{-2},\ \Lambda
=10^{-56}cm^{-2}.  \label{InflatHub}
\end{equation}

For a positive Cosmological Constant the solutions are asymptotic
inflationary for $K=0,-1.$ For $\Lambda >\Lambda _{crit}$ \ and $K=+1,$ the
scale factor time evolution is similar to the cases that correspond to $%
K=0,-1,$ where eventually the asympotic de-Sitter Universe is reached.

\begin{figure}
\epsfxsize=6.5in
\epsfysize=7.5in
\epsffile{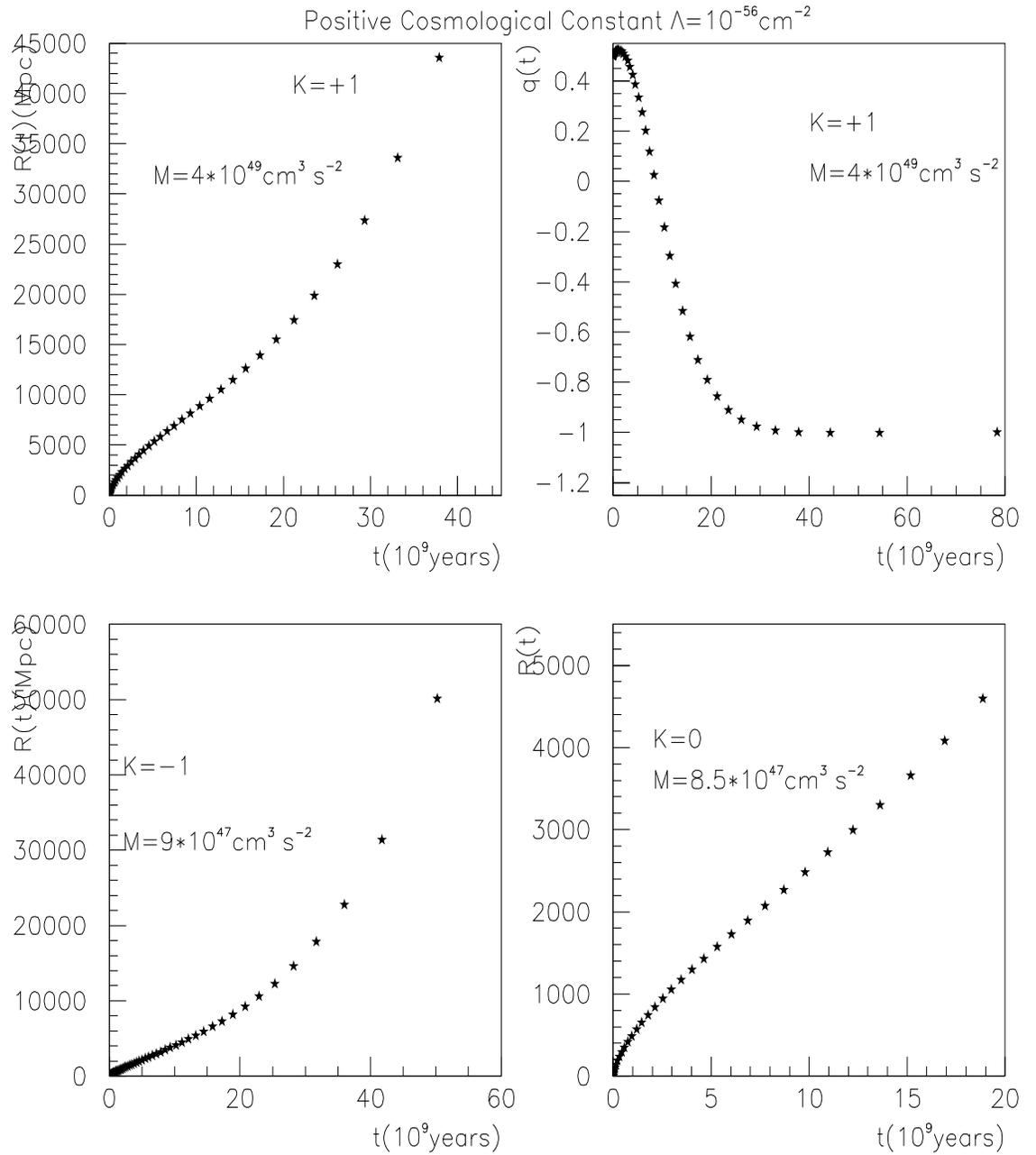}
\caption{Asymptotic inflationary
cosmological scenarios for positive Cosmological Constant.}
\label{ASYMPTOTICDESITTER}
\end{figure}

Interestingly for the case, $K=+1$ and $0<\Lambda <\Lambda _{crit}$, a
periodic or a bouncing solution is obtained. This class of solutions
includes the special case $g_{3}=0,g_{2}\not=0$ discussed earlier in the
paper, where $\Lambda =\frac{1}{2}\Lambda _{crit}$ (see also next section).
The resulting solutions are shown in fig.(\ref{ASYMPTOTICDESITTER}).
Again, the solutions show a general trend. Initially matter dominates the
evolution of the scale factor, it then goes through the point of inflection
and finally tends asymptotically to infinity. The Hubble's Constant in this
last phase is determined entirely by the Cosmological Constant Eq.(\ref%
{Asymptotic}) whose asymptotic value is $H_{\infty }^{2}=\left( \frac{%
c^{2}\Lambda }{3}\right) \cong \left( 54\frac{Km}{sMpc}\right) ^{2},$ for
the value of $\Lambda $\ chosen in Eq.(\ref{InflatHub}).

Similarly, we observe that the deceleration parameter starts with a positive
value ($q=0.5)$,  goes through zero at the inflection points and tends
asymptotically to minus one. For this choice of the cosmological parameters,
a ''good'' set of results, for $K=0$, is given by:

\begin{tabular}{ll}
Scale Factor & 3300 Mpc \\ 
Age of Universe & 13.6 Billion Years \\ 
Hubbles Constant & 66 $Km\;s^{-1}Mpc^{-1}$ \\ 
Deceleration Parameter & -0.477%
\end{tabular}

\bigskip

The above cosmological values are in agreement with recently published
experimental results: scale factor \cite{OHANIAN}, deceleration parameter $q$
\cite{PERLMUTTER} and Hubble's Constant \cite{PERLMUTTER}\cite{TURNER}.

\subsubsection{Periodic (special case) and Bouncing Universe}

This section examines the class of models that lead to the periodic and
bouncing Universe. These solutions are produced when, $K=1$ and $0<\Lambda
<\Lambda _{Crit}$. For this case we have chosen a parameter set given by,

\begin{equation}
K=+1,\quad M=9\times 10^{47}cm^{2}s^{-2},\quad \Lambda =10^{-56}cm^{-2}.
\end{equation}

\begin{figure}
\epsfxsize=6.5in
\epsfysize=7.5in
\epsffile{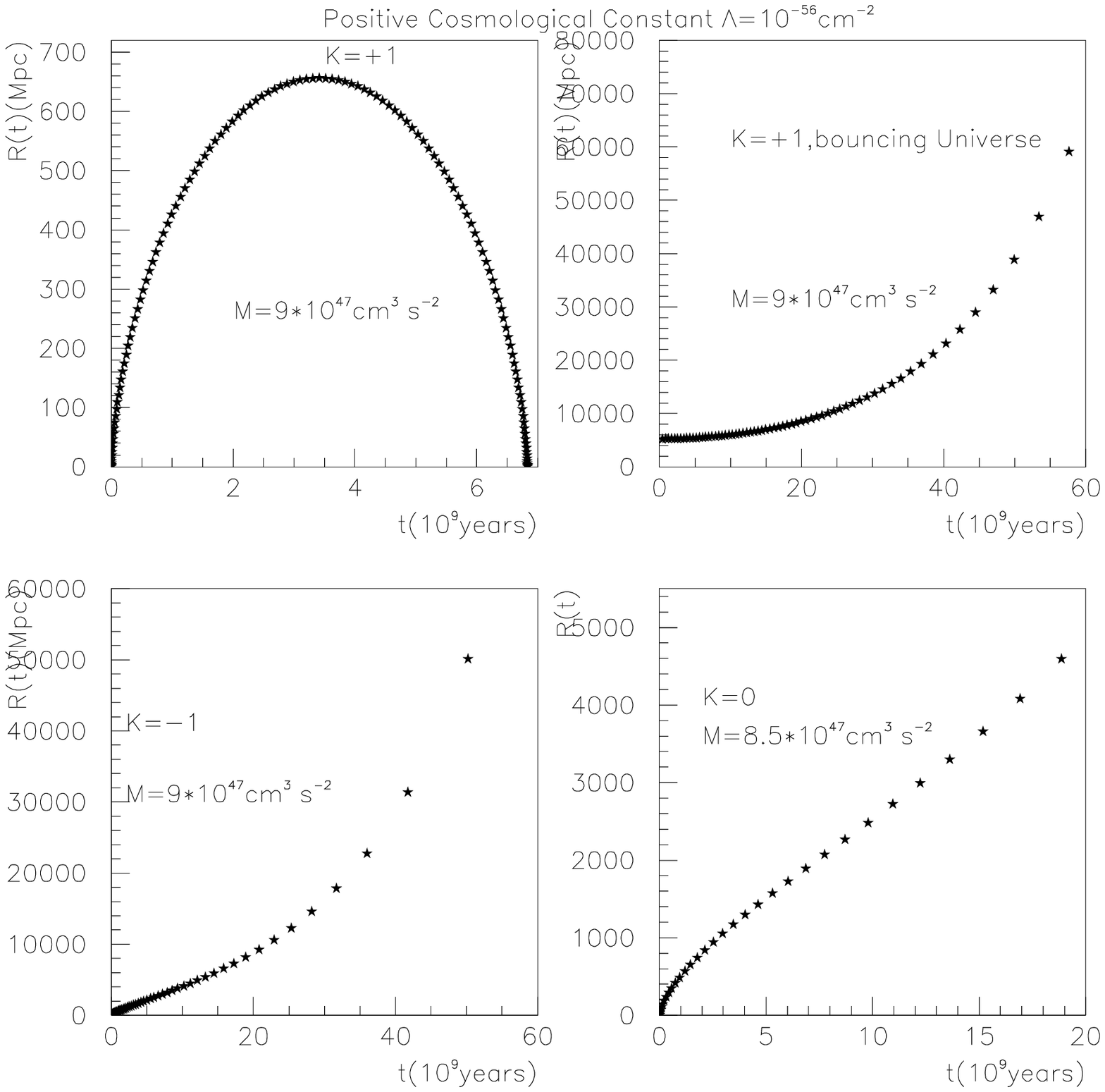}
\caption{Scale factor vs time for $%
0<\Lambda <\Lambda _{crit}.$}
\label{LAMBDAGLCRIT}
\end{figure}

The resulting solutions are shown in the first two 
graphs of fig.(\ref{LAMBDAGLCRIT}). For comparison the cases with $K=-1,0$ and same value for the $\Lambda, M$ parameters are shown in the last two graphs of 
fig.(\ref{LAMBDAGLCRIT}).

\bigskip

The bouncing solution is of particular interest in that there is no $R=0$
singularity, and the periodic solution, for this special case, is similar to
solutions with negative Cosmological Constant and positive curvature
coefficient.


\bigskip



\bigskip

\bigskip

\bigskip

\bigskip

\bigskip

\bigskip

\bigskip \newpage

\bigskip

\section{Discussion}

A new exact solution of the field equations of General Relativity, in closed form in terms of the Weierstra$\ss$ modular form for the Szekeres-Szafron family of
inhomogeneous space time line element with a non-zero Cosmological Constant
has been derived. Our approach  was to integrate the non-linear PDEs by bringing them 
to the form of the defining equation of the elliptic curve by appropriate 
substitutions.
By careful choice of parameters and constants the homogeneous
FLRW limit of the space time line element was recovered. 
The same techniques used for the integration of the field equations 
resulting from the  Szekeres-Szafron spacetime  were applied for the 
integration of the Friedmann field equation. 
Within the
homogeneous cosmology paradigm a useful, predictive theory of cosmology has
been developed. Here knowledge of the cosmological parameters $K,$ $\Lambda $
and $M$ leads to the derivation of the geometry of the Universe where actual
values for the scale factor, Hubble's Constant, deceleration parameter 
are obtained.

For the cosmological parameters, $K=0,$ $\Lambda =+10^{-56}cm^{-2},$ $%
M=8.5\times 10^{47}cm^{3}s^{-2}$, a set of values for the cosmological 
quantities are shown below:

\begin{table}[h]
\begin{center}
\begin{tabular}{|c||c|}  \hline\hline
Scale factor $R(t)$ & $3300\;Mpc$ \\
Age of the Universe & $13.6$ Billion years \\
Deceleration parameter & $-0.477$ \\
Hubble's Constant & 66${\rm Km s^{-1} Mpc^{-1}}$, \\
 \hline\hline
\end{tabular}
\end{center}
\end{table}

which are in reasonable agreement with the published experimental values %
\cite{TURNER},\cite{PERLMUTTER}.

However a set of cosmological parameters with a negative Cosmological
Constant also predict a ''reasonable'' set of values, the essential
difference between the two is the sign and magnitude of the deceleration
parameter ($q=-0.48$ for $\Lambda >0,$ $q=2.2$ for $\Lambda <0).$ It will be
interesting to see how the value for the deceleration parameter changes over
time and consequently the geometry of the Universe, as more experimental
data become available.

The body of the paper shows how the use of Weierstra$\ss$ functions, and the
associated theory of elliptic curves and modular forms, provides the
mathematical framework for solving several non-linear equations arising from
General Relativity. Further we suggest (see appendix D) that many other non-linear 
partial differential equations of mathematical physics 
could be solved from just such an approach. 
There we give examples from soliton physics.
Thus the methodology and the underlying philosophy developed in 
this paper is 
quite generic and suggests a $unified\;approach$ for investigating and solving various physical 
problems of mathematical physics.

Weierstra$\ss$ functions are important in that they provide the link between
pure mathematics analysis techniques and the physical problems of
theoretical cosmology. \ In particular, for a positive Cosmological Constant 
$\Lambda $ and $K=0$ (Euclidean Universe), the discriminant $\Delta $ is
negative resulting in one negative real root. In this case the integration
path includes a zero point, $v_{0},$ of the Weierstra\ss\ function ($\wp
(v_{0})=0)$ which invariably leads to an asymptotically de Sitter phase in
which the Cosmological Constant dominates the dynamics and the fate 
of the Universe, i.e. 
$\frac{c^{2}\Lambda }{3}=H_{\infty }^{2}.$The zeros of the $\wp $ function
in this case are points of order $3$ (i.e. they are points of the FPP that 
satisfy $3P=%
\mathcal{O}$) and belong to the torsion subgroup of the elliptic curve
group. 
The eternally accelerating universe (which will exhibit an {\em event horizon})
is associated with the zeros of a modular form!
In the general case of positive Cosmological Constant, and for $K=-1$
and $K=+1$ ($\Lambda >\Lambda _{crit})$, \ we also reach the dynamical de
-Sitter phase asymptotically at the points $v_{0}$ of the fundamental domain
which solve the equation $\wp (v_{0})=-\frac{K^{\prime }}{12}.$
We also note that in the special case $g_3=0,g_2 \not =0$ the Universe recollapses even for a positive Cosmological Constant for $K=+1$.

For a negative Cosmological Constant and $\Delta <0$ , the Weierstra\ss\
function is always positive in the physical region of integration in the
FPP, \ and its values lie between the positive real root and $+\infty .$ In
this case the integration path in the FPP never includes the points that
solve the equation $\wp (v_{0})=-\frac{K^{\prime }}{12}$ and a periodic
solution is always obtained. The maximum scale factor occurs at the $2-$%
division points, $e_{2}$ =$\wp (\omega +\omega ^{\prime })$ of the
Weierstra\ss\ function. These points are points of order $2$ which also
belong to the torsion subgroup of the elliptic curve group. The value of the
positive real root determines the maximum value for the scale factor.
Similar arguments hold for the positive discriminant case.

Oscillating universes have a certain poetic almost transcedental appeal and many ancient cultures fostered the notion of a universe that periodically died, only to rise phoenix-like from the ashes to seed a new creation. As described by Herodotus \cite{HERODOTUS}: ``the sacred bird was indeed a 
great rarity, even in Egypt, 
only coming there according to the accounts of the people of Heliopolis 
once in five hundred years when the old phoenix dies.'' For a recent account ofthe oscillating case see the book by Barrow \cite{JOHN}.  

The theory of Taniyama-Shimura \cite{GORO,Taniyama} has interesting ramifications, for the
special cases of the invariants $g_{2}=0,g_{3}\not=0$ and $g_{2}\not=0$,$%
g_{3}=0,$ which correspond to the Euclidean and $\Lambda $=$\frac{1}{2}%
\Lambda _{crit}$ solutions. These solutions correspond to elliptic curves
with the property of complex multiplication and are connected by the
Taniyama-Shimura theory, to modular elliptic curves. 
The corresponding tori admit extra symmetries and their complex structure 
belongs to a quadratic imaginary field (in fact the complex structure is 
a fixed point of the modular group).
Consequently,  the
solutions are intimatetly connected with the hyperbolic upper half-complex
plane. For discussions in the mathematical literature of the property 
of complex multiplication we refer the reader to \cite{GORO,Taniyama}.

We believe that the machinery developed in this paper based on exact solutions of General Relativity should be a useful tool 
for observational cosmology and 
for investigating in 
a precise manner the physical implications and 
mathematical properties of various theoretical models. 
Finally, the presently accepted model of the Universe corresponds to a
solution that admitts complex multiplication ($g_2=0,g_3 \not =0$) and therefore has maximal
symmetry. We believe that this is not coincidential but points to a deep
connection between General Relativity and topology. This raises the question
of Einstein:``Did God had any choice in creating the Universe?''.

\newpage

\appendix

\section{\protect\bigskip Ricci Tensor and Scalar for the Szekeres-Szafron
Metric}

Below are the equations for the non-zero components of the Ricci tensor, $%
R_{\mu \nu }$ and \ $R$ scalar.

\begin{equation}
R_{00}=2\frac{\partial ^{2}\beta }{\partial t^{2}}+\frac{\partial ^{2}\alpha 
}{\partial t^{2}}+2\left( \frac{\partial \beta }{\partial t}\right)
^{2}+\left( \frac{\partial \alpha }{\partial t}\right) ^{2}.
\end{equation}

\begin{eqnarray}
R_{11} &=&-e^{2\beta }\left[ \frac{\partial ^{2}\beta }{\partial t^{2}}%
+2\left( \frac{\partial \beta }{\partial t}\right) ^{2}+\frac{\partial \beta 
}{\partial t}\frac{\partial \alpha }{\partial t}\right]  \notag \\
&&+\frac{\partial ^{2}\beta }{\partial x^{2}}+\frac{\partial ^{2}\alpha }{%
\partial x^{2}}+\left( \frac{\partial \alpha }{\partial x}\right) ^{2}-\frac{%
\partial \beta }{\partial x}\frac{\partial \alpha }{\partial x}+\frac{%
\partial ^{2}\beta }{\partial y^{2}}+\frac{\partial \beta }{\partial y}\frac{%
\partial \alpha }{\partial y}  \notag \\
&&+e^{2(\beta -\alpha )}\left[ \frac{\partial ^{2}\beta }{\partial z^{2}}%
+2\left( \frac{\partial \beta }{\partial z}\right) ^{2}-\frac{\partial
\alpha }{\partial z}\frac{\partial \beta }{\partial z}\right]
\end{eqnarray}

\begin{eqnarray}
R_{22} &=&-e^{2\beta }\left( \frac{\partial ^{2}\beta }{\partial t^{2}}%
+2\left( \frac{\partial \beta }{\partial t}\right) ^{2}+\frac{\partial \beta 
}{\partial t}\frac{\partial \alpha }{\partial t}\right) +\frac{\partial
^{2}\beta }{\partial x^{2}}+\frac{\partial \alpha }{\partial x}\frac{%
\partial \beta }{\partial x}  \notag \\
&&+\frac{\partial ^{2}\beta }{\partial y^{2}}+\frac{\partial ^{2}\alpha }{%
\partial y^{2}}-\frac{\partial \beta }{\partial y}\frac{\partial \alpha }{%
\partial y}+\left( \frac{\partial \alpha }{\partial y}\right) ^{2}  \notag \\
&&+e^{2(\beta -\alpha )}\left( \frac{\partial ^{2}\beta }{\partial z^{2}}%
+2\left( \frac{\partial \beta }{\partial z}\right) ^{2}-\frac{\partial
\alpha }{\partial z}\frac{\partial \beta }{\partial z}\right)
\end{eqnarray}

\begin{eqnarray}
R_{33} &=&-e^{2\alpha }\left( \frac{\partial ^{2}\alpha }{\partial t^{2}}%
+\left( \frac{\partial \alpha }{\partial t}\right) ^{2}+2\frac{\partial
\alpha }{\partial t}\frac{\partial \beta }{\partial t}\right) +e^{2(\alpha
-\beta )}\left( \frac{\partial ^{2}\alpha }{\partial x^{2}}+\left( \frac{%
\partial \alpha }{\partial x}\right) ^{2}\right)  \notag \\
&&+e^{2(\alpha -\beta )}\left( \frac{\partial ^{2}\alpha }{\partial y^{2}}%
+\left( \frac{\partial \alpha }{\partial y}\right) ^{2}\right)  \notag \\
&&+2\left( \frac{\partial ^{2}\beta }{\partial z^{2}}+\left( \frac{\partial
\beta }{\partial z}\right) ^{2}-\frac{\partial \alpha }{\partial z}\frac{%
\partial \beta }{\partial z}\right)
\end{eqnarray}

\begin{eqnarray}
R &=&4\frac{\partial ^{2}\beta }{\partial t^{2}}+6\left( \frac{\partial
\beta }{\partial t}\right) ^{2}+4\frac{\partial \beta }{\partial t}\frac{%
\partial \alpha }{\partial t}+2\left( \frac{\partial \alpha }{\partial t}%
\right) ^{2}+2\frac{\partial ^{2}\alpha }{\partial t^{2}}  \notag \\
&&-2e^{-2\beta }\left( \frac{\partial ^{2}\beta }{\partial x^{2}}+\frac{%
\partial ^{2}\alpha }{\partial x^{2}}+\left( \frac{\partial \alpha }{%
\partial x}\right) ^{2}\right)  \notag \\
&&-2e^{-2\beta }\left( \frac{\partial ^{2}\beta }{\partial y^{2}}+\frac{%
\partial ^{2}\alpha }{\partial y^{2}}+\left( \frac{\partial \alpha }{%
\partial y}\right) ^{2}\right)  \notag \\
&&-e^{-2\alpha }\left( 4\frac{\partial ^{2}\beta }{\partial z^{2}}+6\left( 
\frac{\partial \beta }{\partial z}\right) ^{2}-4\frac{\partial \alpha }{%
\partial z}\frac{\partial \beta }{\partial z}\right)
\end{eqnarray}

\section{Integration of the Non-Linear PDE}

The elliptic integral for time can be defined as:

\begin{eqnarray}
t &=&\int [-K(z)+2M(z)\Phi ^{-1}+\frac{\Lambda }{3}\Phi ^{2}]^{-1/2}d\Phi , 
\notag \\
&=&\int \left\{ \Phi ^{-1}\left[ -K(z)\Phi +2M(z)+\frac{\Lambda }{3}\Phi ^{3}%
\right] \right\} ^{-1/2}d\Phi ,  \notag \\
&=&\int \Phi ^{1/2}\left[ -K(z)\Phi +2M(z)+\frac{\Lambda }{3}\Phi ^{3}\right]
^{-1/2}d\Phi .
\end{eqnarray}

let us first consider the integral,

\begin{equation}
u=\int \Phi ^{-1/2}\left[ -K(z)\Phi +2M(z)+\frac{\Lambda }{3}\Phi ^{3}\right]
^{-1/2}d\Phi .
\end{equation}

Defining a new variable $X=-\frac{1}{\Phi }$ enables the integral to be
reduced in cubic form, given by:

\begin{equation}
u=\int \left[ -2M(z)X^{3}-K(z)X^{2}+\frac{\Lambda }{3}\right] ^{-1/2}dX
\end{equation}

Now by defining a second variable $X=-\frac{\xi +K(z)/12}{M(z)/2},$ the
above integral can be written in a Weierstra\ss\ normal form:

\begin{equation}
u=\int \left( 4\xi ^{3}-\frac{K^{2}(z)}{12}\xi -\frac{1}{216}K^{3}(z)+\frac{%
\Lambda M^{2}(z)}{12}\right) ^{-1/2}d\xi ,
\end{equation}

Thus $\xi =\wp (u+\epsilon ),$ and $\epsilon $ is a constant of integration.
The cubic invariants are:

\begin{equation}
g_{2}=K^{2}(z)/12,\qquad g_{3}=\frac{1}{216}K^{3}(z)-\frac{\Lambda }{12}%
M^{2}(z)
\end{equation}

Finally, substituting back into our previous equations, gives:

\begin{equation}
\Phi =-1/X=\frac{M(z)/2}{\wp (u+\epsilon )+K(z)/12}
\end{equation}

and 
\begin{equation}
t=\int \Phi du=\int \frac{M(z)/2}{\wp (u+\epsilon )+\frac{K(z)}{12}}du.
\end{equation}

\section{Derivation for the Energy Density expression}

In this appendix we derive the expression for the energy density using the
first field equation \ref{EINSTEIN1}, given by,

\begin{equation}
G_{0}^{0}-2G_{\xi }^{\xi }-G_{3}^{3}=+2\ (\ddot{\alpha}+2\ddot{\beta}+\dot{%
\alpha}^{2}+2\dot{\beta}^{2})=-\kappa \left( \rho +3P\right) ,
\end{equation}

\begin{eqnarray}
G_{0}^{0}-2G_{\xi }^{\xi }-G_{3}^{3} &=&2\left[ \frac{\Phi (\Phi
_{,tt}^{\prime }+\nu ^{\prime }\Phi _{,tt})}{\Phi (\Phi ^{\prime }+\Phi \nu
^{\prime })}+\frac{2\Phi _{,tt}}{\Phi }\frac{(\Phi ^{\prime }+\Phi \nu
^{\prime })}{(\Phi ^{\prime }+\Phi \nu ^{\prime })}\right]  \notag \\
&=&\frac{2}{(\Phi ^{\prime }+\Phi \nu ^{\prime })}\left[ \Phi
_{,tt}^{\prime }+\nu ^{\prime }\Phi _{,tt}+2\frac{\Phi _{,tt}}{\Phi }(\Phi
^{\prime }+\Phi \nu ^{\prime })\right]  \notag \\
&=&2e^{\nu }\left[ e_{,z}^{\beta }\right] ^{-1}\{\Phi _{,tt}^{\prime }+\nu
^{\prime }\Phi _{,tt}+2\frac{\Phi _{,tt}}{\Phi }(\Phi ^{\prime }+\Phi \nu
^{\prime })\}  \label{INTERMEDIATE}
\end{eqnarray}

Expression (\ref{INTERMEDIATE}) can be written as follows:

\begin{eqnarray}
\Phi _{,tt}^{\prime }+\nu ^{\prime }\Phi _{,tt}+2\frac{\Phi _{,tt}}{\Phi }%
(\Phi ^{\prime }+\Phi \nu ^{\prime }) &=&  \notag \\
&&\frac{2M(z)}{\Phi ^{3}}\Phi _{,z}-\frac{1}{\Phi ^{2}}M_{,z}(z)-\frac{1}{3}%
\frac{\kappa \Phi _{,z}}{\Phi ^{3}}\int P\left( \frac{\partial \Phi ^{3}}{%
\partial t}\right) dt-\frac{\kappa P}{2}\Phi _{,z}  \notag \\
&&+\frac{1}{6}\frac{\kappa }{\Phi ^{2}}\int P\left( \frac{\partial ^{2}\Phi
^{3}}{\partial t\partial z}\right) dt-\nu _{,z}\frac{M(z)}{\Phi ^{2}}+\nu
_{,z}\frac{\kappa }{6\Phi ^{2}}\int P\left( \frac{\partial \Phi ^{3}}{%
\partial t}\right) dt  \notag \\
&&-\frac{\kappa P}{2}\Phi _{,z}+\left[ \frac{-2M(z)}{\Phi ^{3}}+\frac{1}{3}%
\frac{\kappa }{\Phi ^{3}}\int P\left( \frac{\partial \Phi ^{3}}{\partial t}%
\right) dt-\kappa P\right] (\Phi _{,z}+\Phi \nu _{,z})  \notag \\
&=&\frac{-M_{,z}(z)}{\Phi ^{2}}+\frac{1}{6}\frac{\kappa }{\Phi ^{2}}\int
P\left( \frac{\partial ^{2}\Phi ^{3}}{\partial t\partial z}\right) dt-\frac{%
3M(z)\nu _{,z}}{\Phi ^{2}}  \notag \\
&&+\frac{1}{2}\frac{\kappa \nu _{,z}}{\Phi ^{2}}\int P\left( \frac{\partial
\Phi ^{3}}{\partial t}\right) dt  \notag \\
&&-\frac{3\kappa P}{2}(\Phi _{,z}+\Phi \nu _{,z})  \label{CLOSEBY}
\end{eqnarray}

Eq.(\ref{INTERMEDIATE}) can be written using (\ref{CLOSEBY}),

\begin{eqnarray}
& &G_{0}^{0}-2G_{\xi }^{\xi }-G_{3}^{3} 
= - \notag 
\lbrack \frac{2M_{,z}(z)}{\Phi ^{2}}-\frac{1}{3}\frac{\kappa }{\Phi ^{2}}%
\int P\left( \frac{\partial ^{2}\Phi ^{3}}{\partial t\partial z}\right) dt+%
\frac{6M(z)\nu _{,z}}{\Phi ^{2}}-\frac{\kappa }{\Phi ^{2}}\nu _{,z}\int
P\left( \frac{\partial \Phi ^{3}}{\partial t}\right) dt+ \notag \\
&+& 3\kappa P(\Phi _{,z}+\Phi \nu _{,z})]  \notag 
\times \left\{ e^{\nu }[e_{,z}^{\beta }]^{-1}\right\}  \notag \\
&=&-\kappa (\rho +3P)
\end{eqnarray}%
Finally we get,

\begin{equation}
\kappa \rho =e^{\nu }[e_{,z}^{\beta }]^{-1}\left\{ \frac{2M_{,z}(z)}{\Phi
^{2}}-\frac{1}{3}\frac{\kappa }{\Phi ^{2}}\int P\left( \frac{\partial
^{2}\Phi ^{3}}{\partial t\partial z}\right) dt+\frac{6M(z)\nu _{,z}}{\Phi
^{2}}-\frac{\kappa }{\Phi ^{2}}\nu _{,z}\int P\left( \frac{\partial \Phi ^{3}%
}{\partial t}\right) dt\right\}
\end{equation}

\section{Differential equation of the $\wp$ function, and non-linear equations
  of physics}

The Weierstra\ss\ $\wp(u)=\wp(u,\tau)=\wp(u|\omega,\omega^{\prime})$ 
is an even meromorphic elliptic function of periods $2\omega,2\omega^{\prime}$ 
(i.e. $\wp(u+2\omega)=\wp(u)=\wp(u+2\omega^{\prime})$ for all complex 
numbers $u$), which is of order 2, has a double pole at $u=0$ and the principal part of the function at this pole being $u^{-2}$.   
Thus the Weierstra\ss\ $\wp $ function, for $\tau \in \Im $ (upper half-complex
plane) and $u\in C$ is defined as follows:
\begin{equation}
\wp (u,\tau )=\frac{1}{u^{2}}+\sum_{w\in Z+\tau Z}\left( \frac{1}{(u+w)^{2}}-%
\frac{1}{w^{2}}\right) ,\qquad w\neq 0.
\end{equation}
The Weierstra\ss\ function $\wp $ (as it can
be proven from the definition of the function) has non trivial modular
transformation properties. It is a meromorphic Jacobi modular form of weight
2 and index 0 and signature 1\cite{DON}, namely:

\begin{eqnarray}
\wp \left( \frac{u}{c\tau +d},\frac{a\tau +b}{c\tau +d}\right) &=&(c\tau
+d)^{2}\wp (u,\tau ),\qquad \left( 
\begin{array}{cc}
a & b \\ 
c & d%
\end{array}%
\right) \in SL(2,Z), \\
\wp (u+\lambda \tau +\mu ,\tau ) &=&\wp (u,\tau )\qquad ((\lambda \ \mu )\in
Z^{2}).
\end{eqnarray}

Next we note that the Weierstra\ss\ elliptic function $\wp $
obeys the following important differential equation:

\begin{equation}
(\wp ^{^{\prime }})^{2}=4\wp ^{3}-g_{2}(\tau )\wp -g_{3}(\tau ),
\label{WEIERSTRASS}
\end{equation}%
By differentiating (\ref{WEIERSTRASS}) twice we obtain
\begin{equation}
\wp ^{^{\prime \prime \prime }}(z)=12\wp (z)\wp ^{^{\prime }}(z).
\label{KDV}
\end{equation}

The first equation (\ref{WEIERSTRASS}), is the equation of an
elliptic curve, a Riemann surface of genus 1. The second differential
equation that the $\wp $ function satisfies (\ref{KDV}) is the famous
Kordeweg-de Vries (KdV) non-linear partial differential equation of 
soliton physics (in a time independent form):

\begin{equation}
u_t=u_{xxx}-12uu_x,u=u(x,t)
\end{equation} 

In fact the techniques developed in this paper for solving the gravitational field
equations may have applications to other non-linear differential equations of
mathematical physics. In particular non-linear PDEs of solid state physics
besides (KdV), like \ its two dimensional generalization,
the Kadomtsev-Petviashvili (KP) \cite{SHIOTA}, the Sine-Gordon as well as 
the non-linear
Schr\"{o}dinger\ equation of Quantum mechanics can readily be solved by
elliptic theta functions associated with Riemann surfaces
\cite{ALLIED}. The key property
is that the theta functions associated with a period $\tau $ (or the matrix $%
\Omega $ for a higher genus Riemann surface) that comes from a \ Riemann
surface posses special \emph{function theoretic properties.} One of the most
striking properties is that these special $\vartheta ^{\prime }$s satisfy
simple non-linear partial differential equations of fairly low degree,
similar to the differential equation that the Weierstrass, $\wp ,$ function
obeys.
All the above interesting connections 
have led us to propose a conjecture, namely:
That all non-linear exact solutions 
of the resulting non-linear
partial differential equations (PDEs) of General Relativity  with a
non-zero $\Lambda $ are described by theta
functions $\vartheta (z|\tau ),\vartheta (z|\Omega)$ associated with Riemann surfaces.
Our exact solution given in terms of the 
Weierstra$\ss$ $\wp$ function, of the fairly general class of inhomogeneous 
Szekeres-Szafron metrics already provides strong evidence.
Other  non-linear equations of
General Relativity  whose solutions described by Weierstra\ss\ functions 
have been obtained by (Lemaitre \cite{LEMAITRE}, 
Omer \cite{Omer}, Zecca \cite{ZECCA}).

\section*{Acknowledgements}
It is a pleasure to thank D. Bailin, A. Love and J. Hargreaves for 
discussions and support 
on the work presented in this paper.
We are gratefull to G. Hollyman and D. Whitehouse for their help with 
the graphics and support. GVK is gratefull to the physics department 
of Royal Holloway where most of this work was 
completed, for warm hospitality  during the latter stages of this paper. 
We are also gratefull to the anonymous referees for their very constructive 
comments.
The work of GVK was supported by PPARC.

\bigskip \newpage

\bigskip \newpage

\begin{thebibliography}{99}

\bibitem{ALVERTOS} A. Einstein, ``Kosmologische Betrachtungen zur 
allgemeinen Relativit${\rm \ddot{a}}$tstheorie.'' \emph{
Sitzungsberichte der Preussischen Akademie der Wissenschaften},(1917)142-152

\bibitem{GVKSBW} G. Kraniotis \& S. Whitehouse, \ Proceedings of \emph{%
Sources and detection of Dark Matter and Dark Energy in the Universe, }Los
Angeles, Springer Verlag, Lectures notes in Physics, (2001)66

\bibitem{TURNER} M. Turner, Astro-ph/9904051

\bibitem{COHN} Cohn, Astro-ph/9807128

\bibitem{CARROLL} S. Carroll et. al., Annu. Astron. Astrophys., Vol 30, 499
(1992)

\bibitem{BAHCALL} N.A. Bahcall, \ J.P. Ostriker, S. Perlmutter and P.J.
Steinhardt, Science 284(1999)1481

\bibitem{DEBERNAR} P. deBernardis et al, Nature 404(2000)955; A.H. Jaffe et
al, Phys.Rev.Lett.86(2000)3475

\bibitem{MAEFLP} M.Axenides, E.G. Floratos and L.Perivolaropoulos, Mod.
Phys. Letts. A, Vol. 15, (2000), 1541

\bibitem{PERLMUTTER} S. Perlmutter et al. Astrophys.J. 517(1999),565

\bibitem{FELTENGER} J. Overduin and W. \ Priester, astro-ph/0101484

\bibitem{exact} Exact Solutions of Einstein's Field Equations, D. Kramer, H.
Herlt, M. MacCallum, Cambridge University Press, 1980

\bibitem{KRASINSKI} A. Krasinski, Inhomogeneous Cosmological Models,
Cambridge University Press, 1997

\bibitem{BARROW} J. Barrow \& Stein-Schabes, Phys. Lett. Vol 103A, 315 (1984)

\bibitem{SZAFRON} D. Szafron, J. Math. Phys., Vol 18, No. 8, 1673 (1977)

\bibitem{SZEKERES} P. Szekeres, Commun. Math. Phys., Vol 41, 55 (1975)

\bibitem{WHITAKKER} A Treatise on the Analytical Dynamics of Particles \&
Rigid Bodies, E. Whittaker, Cambridge University Press, 1947

\bibitem{EHLERS} P. Schneider, J. Ehlers, E.E. Falco, Gravitational Lenses, 
Springer (1999)

\bibitem{SA} R.K. Sachs. Proc. Roy. Soc. London, {\bf A264} (1961), 309

\bibitem{Zel} Ya.B. Zel$^{'}$dovich, ``{\it Observations in a universe 
homogeneous in the mean}'', Sov.Astr.{\bf 8}(1964),13

\bibitem{DA651} V.M. Dashevskii $\&$    Ya.B. Zel$^{'}$dovich, 
``{\it The propagation of light in a nonhomogeneous nonflat universe},''
Sov. Astr.,{\bf 8}(1965),854.
 
\bibitem{GU67} J.E. Gunn, Ap.J.,150(1981)737


\bibitem{KANTO} R. Kantowski, ``{\it Corrections in the luminocity-redshift
    relations of the homogeneous Friedmann models.}´´ , Ap.J.,155(1969) 89.

\bibitem{GOODE} Goode \& Weinwright, 1982b, Phys. Rev. D26, 3315

\bibitem{KRASGR} Krasinski, gr-qc/9806039

\bibitem{ABRAMOWITZ} Handbook of Mathematical Functions, Ed. by M.
Abramowitz and I. Stegun, Dover Publications, New York 1965

\bibitem{JOSEPH} J.H. Silverman, J. Tate, Rational Points on Elliptic Curves,
Springer Verlag, (1992)

\bibitem{OHANIAN} H. Ohanian \& R. Ruffini, Gravitation \& Spacetime, Norton
and Company, New York, 1994.

\bibitem{DON} M. Eichler and D. Zagier, The Theory of \ Jacobi Forms, Birkh%
\"{a}user, Progress in Mathematics, Vol.55(1985)

\bibitem{STECKBRIEF} J. Sola, hep-ph/0101134

\bibitem{WILES} A. Wiles, Annals of Mathematics, Vol 142, (1995), 443

\bibitem{TARTACARDA} G. Cardano, \emph{Artis Magnae sive de regvlis
algebraicis. }English translation:\emph{The Great Art or the Rules of
Algebra. }MIT Press, Cambridge, MA, 1968

\bibitem{WEINBE} S. Weinberg, ``Gravitation and Cosmology'', Wiley (1972)


 

\bibitem{Taniyama} Taniyama-Shimura, Tokyo-Nikko Symposium on Algebraic
Number Theory, Sept. 1955

\bibitem{FREY} G. Frey, Links between stable Elliptic Curves and certain
Diophantine Equations, Annals Universitatis Saraviensis Vol 1, 1 (1986)

\bibitem{JPSERRE} J-P. Serre, \emph{Sur les repr\'{e}sentations modulaires
de degr\'{e} 2 de }Gal($\bar{Q}/Q$), Duke Math.J. 54(1987) no.1,179-230

\bibitem{RIBET} K. Ribet, On Modular Representations of Gal$(\bar{Q}/Q)$
arising from Modular Forms, Invent. Math. Vol 100, 431 (1990).



\bibitem{LEMAITRE} G. Lemaitre, Ann. Soc. Sci. Bruxelles, Vol A53, 81 (1933)

\bibitem{Omer} G. Omer, Proc. National Acad. of Sci., Vol 53 No. 1, 1 (1965)

\bibitem{ZECCA} A. Zecca, Il Nuovo Cimento, Vol. 106B(1991)N.4


\bibitem{DIAMOND} C. Breuil, B. Conrad, F. Diamond and R. Taylor, \emph{On
the modularity of elliptic curves over Q, or Wild 3-adic exercises, }%
http://www.math.harvard.edu/HTML/Individuals/Richard\_Taylor.html

\bibitem{YUTAKA} G. Shimura, Bull. London Math.Soc.21(1989)186

\bibitem{21} . Harkness, F. Morley, Introduction to the Theory of Analytic
Functions, Macmillan \& Co.,1898

\bibitem{GORO} G. Shimura, Nagoya Math.J. 43(1971)199

\bibitem{SHIOTA} T. Shiota, Inv.Math.83(1986)333-382
\bibitem{ALLIED} I.M.Krichever, Funk.anal.i pril. 11(1977), No 1,15-31
\bibitem{HERODOTUS} Herodotus, {\it The Histories}, Translated by George Rawlinson, Everyman's library,(1992)
\bibitem{JOHN} J.D. Barrow, {\it The Book of nothing} Published by Jonathan Cape (2000); J.D. Barrow and M. Dabrowski, 
Mon.Not.Roy.Astron.Soc.275(1995)850

\end{thebibliography}
\end{document}